\documentclass[lettersize,journal]{IEEEtran}
\usepackage{amsmath,amsfonts}
\usepackage{algorithmic}
\usepackage{array}
\usepackage[caption=false,font=normalsize,labelfont=sf,textfont=sf]{subfig}
\usepackage{textcomp}
\usepackage{stfloats}
\usepackage{url}
\usepackage{verbatim}
\usepackage{graphicx}

\usepackage{cite}
\usepackage[hidelinks]{hyperref}

\usepackage{amsthm}
\usepackage{amsmath} 
\usepackage{orcidlink}

\newtheoremstyle{nonbold} 
  {10pt} 
  {10pt} 
  {\normalfont} 
  {} 
  {\itshape} 
  {:} 
  { } 
  {} 

\theoremstyle{nonbold}

\usepackage{enumitem}



\def\BibTeX{{\rm B\kern-.05em{\sc i\kern-.025em b}\kern-.08em
    T\kern-.1667em\lower.7ex\hbox{E}\kern-.125emX}}
\usepackage{balance}

\begin{document}
\title{Impacts of Physical-Layer Information on Epidemic Spreading in Cyber-Physical Networked Systems }
\author{Xianglai Yuan, Yichao Yao, Han Wu\orcidlink{0000-0001-8870-5694}, $\textit{Member, IEEE}$, Minyu Feng\orcidlink{0000-0001-6772-3017}, $\textit{Senior Member, IEEE}$

\thanks{This work was partially supported by the National Natural Science Foundation of China (NSFC) under Grant No. 62206230, and partially by the Natural Science Foundation of Chongqing under Grant No. CSTB2023NSCQ-MSX0064. \textit{(Corresponding author: Minyu Feng.)}

Xianglai Yuan, Yichao Yao, and Minyu Feng are with the College of Artificial Intelligence, Southwest University, Chongqing 400715, China (e-mail: myfeng@swu.edu.cn).

Han Wu is with the School of Electronics \& Computer Science, University of Southampton, Southampton SO17 1BJ, United Kingdom.
}}

\markboth{IEEE TRANSACTIONS ON CIRCUITS AND SYSTEMS—I: REGULAR PAPERS}
{How to Use the IEEEtran \LaTeX \ Templates}

\maketitle

\begin{abstract}
Since Granell et al. proposed a multiplex network for information and epidemic propagation, researchers have explored how information propagation affects epidemic dynamics. However, the role of individuals acquiring information through physical interactions has received relatively less attention. In this work, we introduce a novel source of information: physical-layer information, and derive the epidemic outbreak threshold using the Microscopic Markov Chain Approach (MMCA). Our simulation results indicate that the outbreak threshold derived from the MMCA is consistent with the Monte Carlo (MC) simulation results, thereby confirming the accuracy of the theoretical model. Furthermore, we find that the physical-layer information effectively increases the population's awareness density and the infection threshold \(\beta_c\), while reducing the population's infection density, thereby suppressing the spreading of the epidemic. Another interesting finding is that when the density of 2-simplex information is relatively high, the 2-simplex plays a role similar to pairwise interaction, significantly enhancing the population's awareness density and effectively preventing large-scale epidemic outbreaks. In addition, our model works equally well for cyber-physical systems with similar interaction mechanisms, while we simulate and validate it in a real grid system.

\end{abstract}


\section{Introduction}\label{intro}
\IEEEPARstart{S}{ince} Duncan J. Watts et al. proposed the WS small-world network \cite{r1} and Albert-László Barabási et al. introduced the BA scale-free network \cite{r2}, research on real-world networked systems have often been abstracted as research on complex networks, where nodes represent entities in real-world networks, and edges represent interactions between these entities \cite{r3,r4}. Some systems consist of multiple complex networks intertwined in complex relationships, often interdependent and inseparable \cite{r5,r6}. For example, in communication and power systems, the communication system requires power from the power system to function, while the power system relies on information from the communication system to distribute power resources appropriately \cite{r7}. Systems with similar dependencies include social networks, transportation systems, and ecosystems \cite{r8, r9}. Meanwhile, research on information propagation and epidemic spreading has also garnered increasing attention \cite{r10,r11, brodka}.

In recent years, Li et al. have refined theoretical propagation models by considering the birth-death process of nodes and introducing the concept of self-protection level, which describes an individual's awareness of self-protection \cite{lyhbhd, feng}. Funk et al., in their work on single-layer networks, explored the coupled propagation of information and epidemics, finding that in a well-mixed population, information diffusion can lead to a lower epidemic outbreak size \cite{funk, ruan}. However, since much information is propagated through virtual networks such as Facebook, Weibo, and TikTok, Granell et al. proposed a multiplex networked system where the upper layer (the virtual contact layer) represents information propagation, and the lower layer (the physical contact layer) represents epidemic spreading \cite{r12}. The nodes in the two layers correspond one-to-one. In the virtual contact layer, nodes can be in one of two states: U (Unaware) or A (Aware), representing whether the node is aware of disease information. In the physical contact layer, nodes can be in one of two states: S (Susceptible) or I (Infected), following the SIS model. The results suggested that effective information propagation can influence the epidemic threshold and reduce infection density \cite{r13, r14}. Bhowmick et al.'s study further indicated that the steady state of an epidemic is not only related to the basic reproduction number \( R_0 \) but is also closely linked to the agents' opinions \cite{tcabc1}, \cite{tcabc2}. Research by Mishkovski et al. showed that the epidemic outbreak threshold in a multiplex network can be higher compared with the isolated networks \cite{bc1}. Feng et al. further explored epidemic spreading when individuals in an aware state either do not believe in the existence of an epidemic or are aware but neglect preventive measures, and the result suggested that individuals without safety awareness have a minimal impact on the spread of the epidemic when randomly distributed, as the proportion of these individuals increases, the infection scale increases linearly. However, if high-degree nodes are selected, the impact of safety awareness on epidemic spread becomes significant \cite{lxxf}. Guo et al. proposed the LACS model (a local awareness-controlled contagion spreading model), where individuals transition to an aware state and become immune to the epidemic when the proportion of aware neighbors exceeds a certain threshold \cite{guotwo}. Given that interactions frequently occur between multiple nodes in human communication, animal brains, and ecosystems \cite{karsai, cook, mayfield}, Fan et al. introduced a 2-simplex to describe peer influence and reinforcement mechanisms in information propagation, expanding the forms of information propagation \cite{fan}. The research by Zhang et al. suggested that in the two-layer metapopulation network model, self-herd awareness can properly curb the epidemic spreading, and neighboring herd-awareness propagation can decrease the infection scale \cite{zhang2024interplay}. Research by Velasquez-Rojas et al. found that increasing the speed of information propagation can reduce the prevalence of an epidemic\cite{velasquez}. Similarly, Wang et al. showed that in a two-layer metapopulation network, the rate of information propagation can suppress the spreading of the epidemic, however, once the information propagation rate exceeds a certain threshold, its effectiveness in controlling the epidemic diminishes \cite{wang2021impacts}. Gu et al.'s study indicated that unverified information can still help curb the spreading of the epidemic \cite{huo2023influence} while Chen et al. discovered that negative information generated during vaccination can lead to an increase in infection density, and an individual's willingness to vaccinate plays a dominant role in influencing infection density \cite{chen2022coevolving}.

While the above models enrich the information propagation pattern, they remain limited to the influence of the information layer on the physical layer and do not fully consider or ignore the ability of physical layer individuals to acquire information and subsequently influence the information diffusion in the information layer. We should recognize the individuals’ agency—their ability to actively perceive information during physical contact and take measures to protect themselves. We know that in the physical layer when individuals interact with others, they typically can initially judge whether someone is infected by talking to them or observing their physical state. When individuals perceive that someone they interact with is infected, they take protective measures to avoid infection. This helps protect themselves, their families, and friends, thereby preventing the spreading of the epidemic. The Health belief model in psychology suggests that individuals who perceive a specific health issue as serious are more likely to take preventive actions to avoid the infection \cite{HBM}. Additionally, Kahneman et al. suggest that when an individual’s perception reaches a specific critical point or threshold, their nature, state, or behavior undergoes a significant change \cite{prospect}. The evolutionary game theory indicates that individuals observe other individuals in the group and adjust their behaviors according to the state of other individuals to maximize their own interests \cite{games}. In epidemiology, Bish et al. highlight the evidence that individuals who perceive themselves as susceptible to diseases such as SARS, avian flu swine flu, or pandemic influenza are more likely to engage in preventive and disease management behaviors to protect themselves \cite{pandemic}.

To reveal the impact of different information on epidemic spreading fully, we construct a novel cyber-physical networked system with a multiplex networked structure. The upper layer denotes the cyber layer, pairwise interactions between nodes indicate direct information propagation between individuals, while the 2-simplex indicates higher-order interactions indicating information propagation shaped by peer influence and reinforcement mechanisms, such as social media, online forums, etc. The lower layer denotes the physical layer, describing the spreading of the epidemic through physical contact in real life. Specifically, to accurately describe the impact of physical-layer information on individuals, we use a nonlinear threshold model to simulate the probability that individuals become aware of the epidemic and take protective measures according to the research above, which is shown in the introduction of 1 - \( r_{i}^{(3)}(t)\). Through extensive simulations, we find that the infection densities derived from the Microscopic Markov Chain Approach (MMCA) closely agree with the ones obtained from Monte Carlo (MC) simulations and the infection density decreases as the information propagation rate increases. Meanwhile, the study reveals that physical-layer information can affect the epidemic outbreak threshold, indicating its potential impact on the transmission dynamics of the entire system.

It is worth noting that some modern cyber-physical systems involving information interactions, such as network security systems, intelligent transportation, and smart grids, can also be studied using this model. For example, we understand that smart grids not only focus on power supply but also involve the management of information flows such as monitoring, data collection, and feedback. The research facilitated by AI in circuits and systems gains a promising development, which in return lead to a surge in electricity demand, making the power system a highly critical infrastructure \cite{bcdw}, \cite{bc11}. Modern cyber-physical systems in the power grid have real-time monitoring, dynamic control, and information services through the seamless integration of computational systems, communication networks, and physical entities \cite{bc2}, \cite{bc3}. However, due to the relatively open nature of communication systems \cite{bc4}, the power system is vulnerable to attacks, which increases the likelihood of power outages and cascading failures, posing significant challenges to the operation and maintenance of the power system \cite{bc5},\cite{bc6}. In smart grids, the transmission of electricity and the dissemination of information resemble the physical layer (power flow) and the cyber layer (control signals, communication information) in a dual-layer network model. Specifically, fault propagation in power grids is analogous to the ``infection" mechanism in epidemic spread. When a node in the grid (e.g., a substation or distribution equipment) experiences a fault, it can trigger a cascading effect, impacting the normal operation of other components or even causing cascading failures. Meanwhile, the grid's communication system can receive and transmit fault information and activate predefined isolation and recovery mechanisms. Based on this, we conduct studies within the U.S. power grid system to study the impact of information propagation on power systems and verify the applicability of the model on real-world networks.

The rest of the content is organized as follows. In Sec. II, we first model the nonlinear cyber-physical networked system and use the MMCA to derive the probability transition equations to analyze the epidemic outbreak threshold. Then, in Sec. III, we compare the results of MC simulations and the MMCA and study the impact of pairwise interaction information, 2-simplex information, and physical-layer information on epidemic spreading and information diffusion based on the constructed nonlinear cyber-physical networked system model, and validate our conclusion through case studies on the U.S. grid further. Finally, in Sec. IV, we provide a comprehensive summary of our findings, discuss the broader implications of our results, and propose potential directions for future research in this area.

\section{Model Description}\label{sec: section 2}

In this section, we first introduce the modeling process of the nonlinear cyber-physical networked system and provide the corresponding theoretical analysis. Then, we display the node state transition process in the model based on MMCA and derive the analytical solution for the epidemic outbreak threshold in the steady state.
\subsection{Modeling of the Cyber-Physical Networked System}
As we attempt to study the coupled dynamics of information propagation and epidemic spreading on a nonlinear cyber-physical networked system, we have the following multiplex network. The upper layer is the cyber layer, while the lower one is the physical layer where the epidemic spreads. To capture the randomness of information spread in the network, the cyber layer is constructed using the Erd\H{o}s--R\'enyi (ER) random graph, considering both pairwise interactions between nodes and additional information diffusion induced by the 2-simplex while the physical layer is built based on the WS network to reflect two important properties of real-world networks: local clustering and short path lengths. The nodes in the upper layer and the nodes in the lower layer are in one-to-one correspondence, which implies that nodes in the network are affected by both information and epidemics. Also, the whole nonlinear cyber-physical networked system is assumed to be undirected and unweighted, as shown in Fig. \ref{fig. 1}. The specific construction process of the cyber layer is as follows:

\begin{figure}[h]
    \centering
    \includegraphics[width=\linewidth]{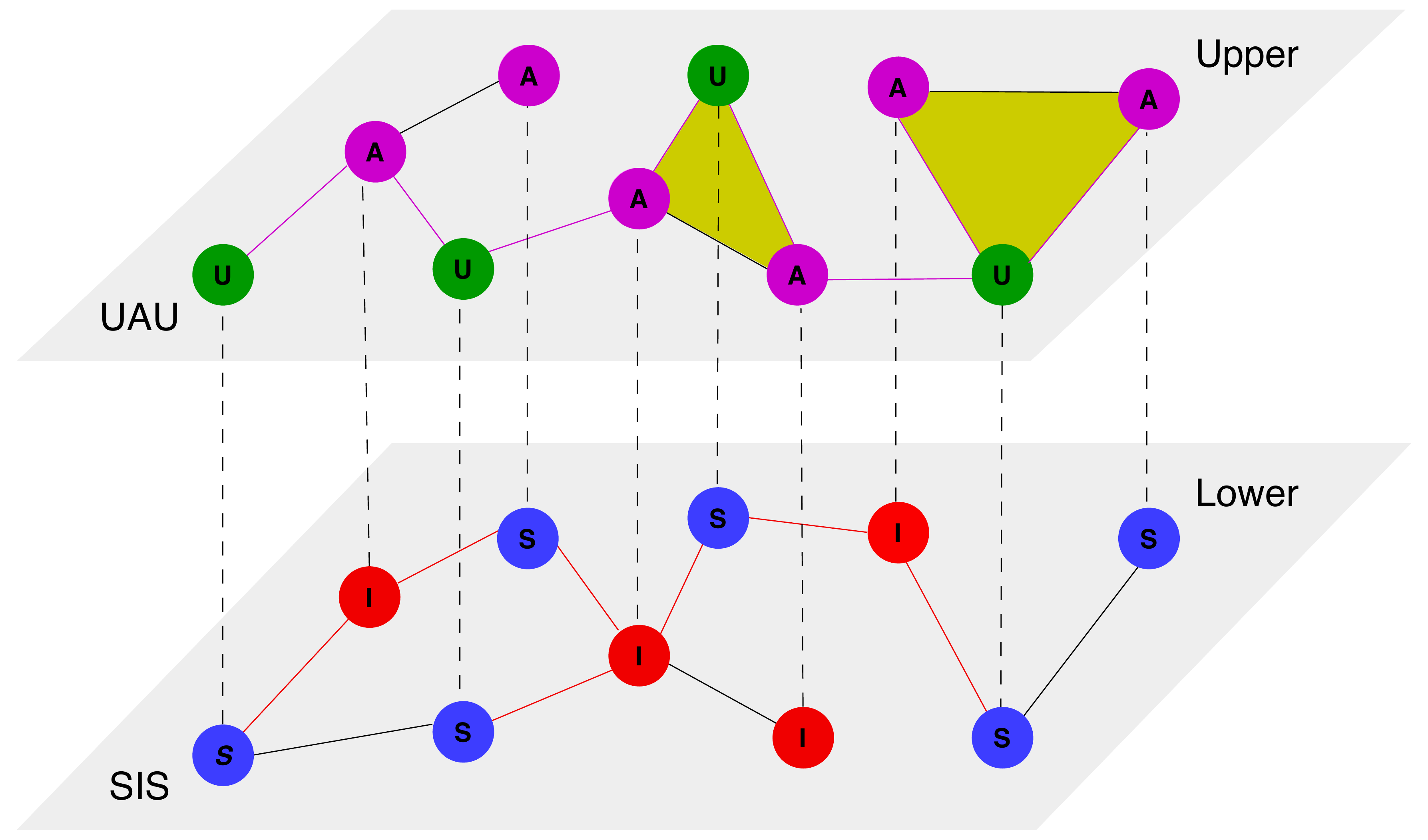}
    \caption{\textbf{The proposed multiplex network.} The upper layer is the cyber layer, simulating the propagation of epidemic information. Nodes can be in one of two states: Unaware (U) or Aware (A). The connections between nodes display the propagation of information, with purple lines indicating that unaware nodes can get informed by the aware nodes. The yellow triangular face denotes the propagation of information through the 2-simplex. The lower layer is the physical layer, describing the spreading of the epidemic, where nodes can be in one of two states: Susceptible (S) or Infected (I). The connections between nodes indicate physical contact, with red lines indicating contact between susceptible and infected nodes while black lines indicate contact between either susceptible or infected nodes. Dashed lines between the two layers indicate one-to-one matching of nodes, and the network itself is undirected and unweighted.}

    \label{fig. 1}
\end{figure}

(i) We generate an ER random graph with \( N \) nodes and a connection probability of \( p_1 \) (\( 0 < p_1 < 1 \)) to denote the cyber layer. The average degree of nodes in the cyber layer is denoted as  \( k_{1} \) while \( k_{2} \) denotes the number of 2-simplexes that each node may locate. Then, at this time, the node's average degree in the cyber layer is \( (N-1)p_1 \).

(ii) Let \( p_2 \) (\( 0 < p_2 < 1 \)) denote the probability of forming a 2-simplex between any three nodes \( i, j, k \) selected from the cyber layer. We generate a 2-simplex with probability  $p_2$  for each set of three nodes in the cyber layer, and the newly formed edges are added to the cyber layer constructed in step (i), updating the original structure. Using the combination formula, it is easy to obtain \( k_2 = {(N-1)(N-2)p_2}/{2}\).

For any node \( i \), we analyze the increase in the degree of node \( i \) caused by the 2-simplex. There are three cases here. First, in the cyber layer constructed in step (i), the likelihood that neither an edge exists between \( i \) and \( j \), nor between \( i \) and \( k \), is \( (1-p_1)^2 \). When both \( j \) and \( k \) are connected to \( i \) to form a 2-simplex, the degree of the node \( i \) increases by 2. Second, the probability of no edge between \( i \) and \( j \), but an edge existing between \( i \) and \( k \), is \( (1-p_1)p_1 \). By connecting \( i \) to \( j \) to form a 2-simplex, the degree of node \( i \) increases by 1. Lastly, the chance of no edge between \( i \) and \( k \), but an edge between \( i \) and \( j \) is \( (1-p_1)p_1 \), after connecting \( i \) to \( j \), the degree of node \( i \) increases by 1. 

To sum up, after constructing the 2-simplex, for any node \( i \) in the network, its degree will increase by \( 2(1-p_1)^2 + 2(1-p_1)p_1 = 2(1-p_1) \). In other words, by constructing the 2-simplex, the network's average degree increases by \( 2(1-p_1) \). Therefore, we have \( k_1 = (N-1)p_1 + 2(1-p_1)k_2 \). By equivalence transformation, we obtain \( p_1 \) and \( p_2 \) as follows:
\begin{equation}\label{eqs. 1}
\left\{
\begin{aligned}
    p_{1} &= \frac{k_1-2k_2}{(N-1)-2k_2}, \\ 
    p_{2} &= \frac{2k_2}{(N-1)(N-2)}.
\end{aligned}
\right.
\end{equation}

\subsection{Analysis of Individual States Based on MMCA}
\begin{figure}[!t]
    \centering
    \includegraphics[width=0.9\linewidth]{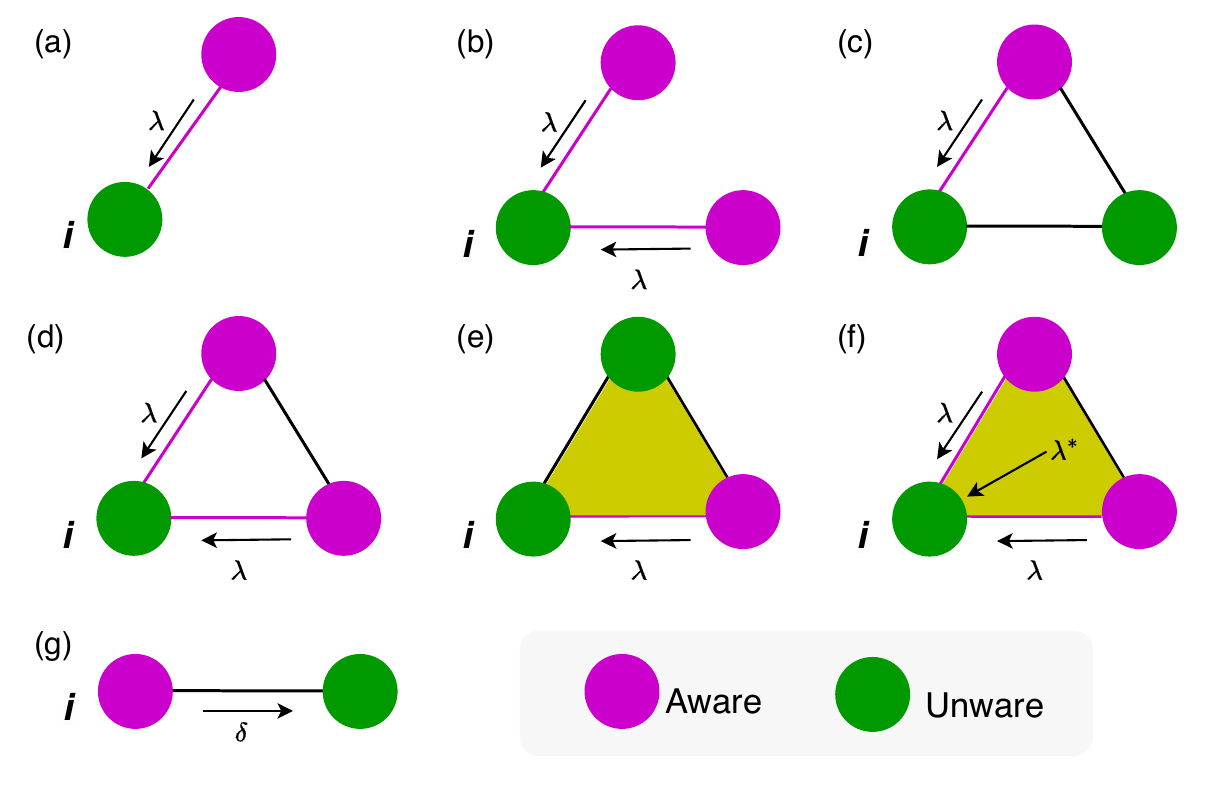}
    \caption{\textbf{Information propagation on the cyber layer.} purple and green nodes indicate nodes in the A state and U state, respectively. In (a)-(d), node i receives information from its neighbors through pairwise interaction (link) with a probability of $\lambda$. In (e), node \( i \) and its two neighbors form a 2-simplex (the triangular enclosure), but since only one neighbor is in the A state, the conditions for information propagation within the 2-simplex are not met. In (f), the 2-simplex is shown where node \( i \) transitions to the A state with probability \( \lambda \) as informed by its A-state neighbor, and under the influence of the 2-simplex, node \( i \) may also become aware with probability \( \lambda^{*} \). Additionally, at each time step, nodes in the A state may revert to the U state due to information loss, with a probability $\delta$, as shown in (g).}
    \label{fig. 2}
\end{figure}

Since individuals may acquire information or lose it, the cyber layer follows the UAU model, where U denotes an unaware state and A denotes an aware state. The transition from U to A occurs for the following reasons: first, by contacting a neighboring node in state A through the pairwise interaction with probability \(\lambda\) or the 2-simplex with probability \(\lambda^{*}\) in the cyber layer, which are shown in Fig. \ref{fig. 2}; second, by being infected or getting the physical-layer information through contact with others in the physical layer. The transition from A to U occurs as the node loses the information with a probability of \(\delta\).

The physical layer of epidemic propagation follows the SIS model to denote the state of an individual, where S denotes an individual in the susceptible state and I indicates an individual in the infected state. A healthy individual can be infected by an infected individual with a probability of \(\beta\), and an infected individual can recover with a probability of \(\mu\). Once an individual is infected, they automatically become aware (A state). To distinguish between aware and unaware nodes, we let \(\beta^U\) be the probability that an unaware susceptible individual is infected, and \(\beta^A\) be the probability that an aware susceptible individual is infected, where \(\beta^A = \gamma \beta^U\) (\(0 \leq \gamma < 1\)) since the nodes in U state are more susceptible to the epidemic than the nodes in A state. Particularly, when \(\gamma = 0\), it means that aware nodes are completely immune to the epidemic.

Based on the above assumption, it is easy to know that a node can be in one of the following states: unaware and susceptible (US), aware and susceptible (AS), and aware and infected (AI). Note that the state of unaware and infected (UI) does not exist, as an individual becomes aware immediately upon infection.

Let \(a_{ij}\) and \(b_{ij}\) represent the elements of the adjacency matrices corresponding to the cyber layer and the physical layer, respectively. \( c_{i} \) denotes the number of 2-simplex around node \( i \), while \( c_{ijk} \) indicates whether nodes \( i \), \( j \), and \( k \) collectively form a 2-simplex. At time step \(t\), the probabilities that node \(i\) is in the states AI (aware-infected), AS (aware-susceptible), and US (unaware-susceptible) are denoted as \(P_{i}^{AI}(t)\), \(P_{i}^{AS}(t)\), and \(P_{i}^{US}(t)\), respectively. Furthermore, let \(r_{i}(t)\) represent the probability that node \(i\) has not received any epidemic-related information from its neighbors. Specifically, \(r_{i}^{(1)}(t)\) denotes the probability that node \(i\) has not been informed through pairwise interaction information, \(r_{i}^{(2)}(t)\) represents the probability that the node has not been notified through 2-simplex information in the cyber layer. 

Notably, Granell et al. were the first to propose a two-layer network propagation model that effectively describes the interaction between the information and physical layers. In their model, they introduced the probability \( k \in [0, 1] \) that an infected individual may remain unaware (UI), despite being infected. Interestingly, they found that this probability \( k \) had minimal impact on the overall epidemic outbreak dynamics \cite{r12}, \cite{r14}. This insight suggests that the presence of unaware infected individuals does not significantly alter the course of the epidemic. Guo et al. introduce a local awareness threshold \( \alpha \) in the information layer, where unaware (U) individuals in the information layer only transition to the aware state (A) when the infection proportion in their vicinity exceeds a certain threshold; otherwise, they remain in the U state \cite{guotwo}. Although those models enrich the information propagation patterns, they remain limited by focusing on the influence of the cyber layer on the physical layer, without adequately considering or neglecting the ability of physical layer individuals to acquire information.

Hence, we let \(r_{i}^{(3)}(t)\) indicate the likelihood that the node has not detected the outbreak by sensing information in the physical layer. Consequently, \(1 - r_{i}^{(3)}(t)\) describes the probability that the individual becomes aware of the epidemic and adopts protective measures through physical-layer information, analogous to \(1 - r_{i}^{(1)}(t)\) and \(1 - r_{i}^{(2)}(t)\) for pairwise interaction and 2-simplex information, respectively. The model of \(1 - r_{i}^{(3)}(t)\) can help effectively capture the nonlinear dynamic response process of individuals based on the infection status of their neighbors. Concretely, \(\theta\) represents the individual's vigilance threshold. If $\theta$ is low ($\theta\to 0^{+}$), the individual is highly sensitive to the epidemic and will adopt protective measures early. Conversely, if \(b\) is high ($\theta\to 1^{-}$), the individual will only consider adopting measures when a large number of people around them are infected. $\alpha$ represents the individual's response intensity to the infection situation. A larger $\alpha$ value indicates that the individual is highly sensitive to changes in infection proportions of neighbors, causing a sharp increase in the probability of adopting protective measures, especially when approaching the vigilance threshold $\theta$. On the other hand, a smaller $\alpha$ value indicates a more gradual response in terms of adopting protective measures.

Lastly, we let \(q_{i}^{A}(t)\) be the probability that node \(i\) is not infected by its neighbors when it is in A state, and \(q_{i}^{U}(t)\) be the probability that node \(i\) is not infected by its neighbors when it is in the U state, then we have:

\begin{equation}\label{eqs. 2}
\left\{
\begin{aligned}
    q_{i}^A(t)  &= \prod_{j}[1-b_{ji}P_{j}^{AI}(t)\beta^A], \\
    q_{i}^U(t) &= \prod_{j}[1-b_{ji}P_{j}^{AI}(t)\beta^U],\\
    r_{i}^{(1)}(t) &= \prod_{j} [1-a_{ji}P_{j}^{A}(t)\lambda], \\
    r_{i}^{(2)}(t) &= \prod_{c_{i}} [1-c_{ijk}P_{j}^{A}(t)P_{k}^{A}(t)\lambda^{*}], \\
    r_{i}^{(3)}(t) &= 1 - \frac{1}{1 + e^{-\alpha (\frac{\sum_{j}b_{ji}P_{j}^{AI}(t)}{k_{i}} - \theta)}}, \\
    r_{i}(t) &= r_{i}^{(1)}(t)r_{i}^{(2)}(t)r_{i}^{(3)}(t),
\end{aligned}
\right.
\end{equation}
where $P_{j}^{A}(t)=P_{j}^{AI}(t)+P_{j}^{AS}(t)$. Thus, we can directly derive the microscopic Markov chain equations for the coupled dynamics of this nonlinear cyber-physical networked system. For each node i, we have:
\begin{equation}\label{eqs. 3}
\left\{
\begin{aligned}
    P_{i}^{US}(t+1) &= P_{i}^{US}(t) r_i(t) q_{i}^{U}(t) 
    + P_{i}^{AS}(t)\delta q_{i}^{U} \\
    &\quad + P_{i}^{AI}(t)\delta\mu, \\
    P_{i}^{AS}(t+1) &= P_{i}^{US}(t) [1 - r_i(t)] q_{i}^{A}(t) \\
    &\quad + P_{i}^{AS}(t) (1 - \delta) q_{i}^{A}(t) \\
    &\quad + P_{i}^{AI}(t) (1 - \delta)\mu, \\
    P_{i}^{AI}(t+1) &= P_{i}^{US}(t)\left[(1 - r_{i}(t)] [1 - q_{i}^{A}(t)] \right] \\
    &\quad + P_{i}^{US}(t) r_{i}(t) [1 - q_{i}^{U}(t)] \\
    &\quad + P_{i}^{AS}(t) \delta [1 - q_{i}^{U}(t)] \\
    &\quad + P_{i}^{AS}(t) (1 - \delta) [1 - q_{i}^{A}(t)] \\
    &\quad + P_{i}^{AI}(t)\left[\delta (1 - \mu) 
    + (1 - \delta) (1 - \mu)\right]\rlap{.}
\end{aligned}
\right.
\end{equation}

\begin{figure*}[!t]
    \centering
    \includegraphics[width=0.9\linewidth]{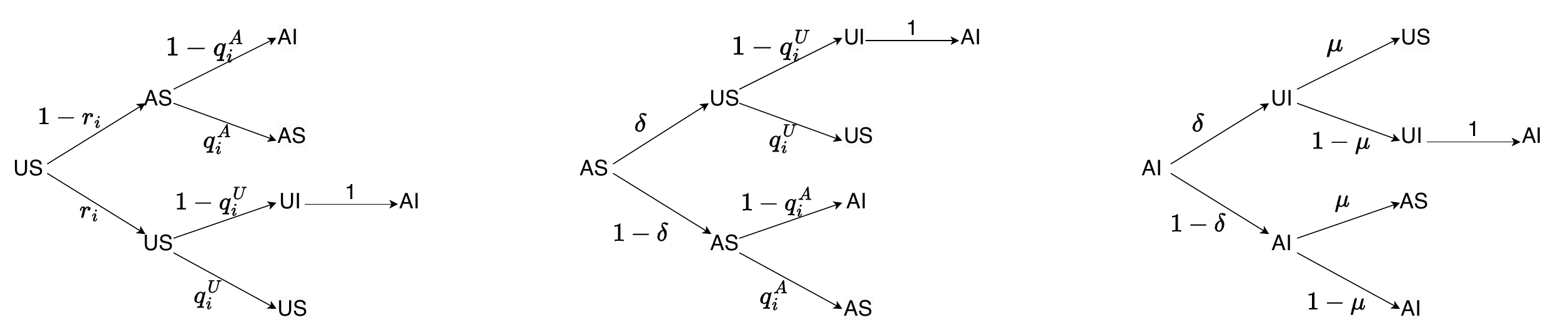}
    \caption{\textbf{Probability tree for transitions among four states (AS, AI, US, UI).} The probability that an individual i in the A state does not get infected is denoted by $q_{i}^{A}$, and the probability that an individual in the U state does not get infected is denoted by $q_{i}^{U}$. $\delta$ denotes the probability that an individual in the A state forgets the information, and $\mu$ denotes the probability that an individual in the I state recovers. It is assumed that when individuals in the UI state contract the epidemic, they immediately transition to the AI state with probability 1.}
    \label{fig. 3}
\end{figure*}

In Eq. \ref{eqs. 3}, for the first equation, the left-hand side is the probability that node \(i\) is in the US state at time \(t+1\); on the right-hand side, the first term is the probability that node \(i\) transitions from the AI state to the US state at time \(t\), the second term is the probability that node \(i\) remains in the US state at time \(t\), and the third term is the probability that node \(i\) transitions from the AS state to the US state at time \(t\). The meanings of the second and third equations in the set are similar, and the transition probability trees of each state are shown in Fig. \ref{fig. 3}.

When the epidemic spreading reaches a steady state, we have $P_{i}^{US}(t+1) = P_{i}^{US}(t) = P_{i}^{US}$, $P_{i}^{AS}(t+1) = P_{i}^{AS}(t) = P_{i}^{AS}$, and $P_{i}^{AI}(t+1) = P_{i}^{AI}(t) = P_{i}^{AI}$. To facilitate the calculation of the epidemic outbreak threshold $\beta_{c}^{U}$, we expand Eq. \ref{eqs. 3} near the critical point. We know that when $\beta^{U}$ approaches the critical value, the epidemic cannot outbreak, and thus the probability of a node being infected is very small, i.e., $P_{i}^{AI} = \epsilon_i$ ($\epsilon_i \to 0^{+}$). Therefore, the second equation in Eq. \ref{eqs. 2}  can be written as $q_{i}^A \approx 1 - \beta^A \sum_{j}b_{ji}\epsilon_j$, and the third equation can be written as $q_{i}^U \approx 1 - \beta^U \sum_{j}b_{ji}\epsilon_j$. Substituting these into Eq. \ref{eqs. 3}, we obtain the following expression:
\begin{equation}\label{eqs. 4}
\left\{
\begin{aligned}
    P_{i}^{US} &= P_{i}^{US} r_{i} + P_{i}^{AS} \delta, \\
    P_{i}^{AS} &= P_{i}^{US}(1-r_{i}) + P_{i}^{AS}(1-\delta), \\
    \mu \epsilon_{i} &= (P_{i}^{AS} \beta^{A} + P_{i}^{US} \beta^{U})\sum_{j} b_{ji}\epsilon_j.
\end{aligned}
\right.
\end{equation}

Since \(P_{i}^{AI} = \epsilon_i \approx 0\) as $\beta_{c}^{U}$ near the point of epidemic outbreak, \(P_{i}^{A} = P_{i}^{AI} + P_{i}^{AS} \approx P_{i}^{AS}\) and $P_{i}^{AS} + P_{i}^{US} \approx 1$. From the second equation in Eq. \ref{eqs. 4}, we conclude that $P^{A}_{i}$ in the critical state is given by:
\begin{equation}\label{eqs. 6}
\begin{aligned}
    P_{i}^{A} &=  (1-\delta)P_{i}^{A} + (1-r_{i})(1 - P_{i}^{A}) \\
\end{aligned},
\end{equation}
where $r_{i}(t) = r_{i}^{(1)}(t) r_{i}^{(2)}(t) r_{i}^{(3)}(t)$, and at this point, $r_{i}^{(3)}(t) \approx 1$. Therefore, Eq. \ref{eqs. 6} can be obtained by solving a system of linear equations or an iterative method using numerical methods. Moreover, the third term in Eq. \ref{eqs. 4} can be further expressed as:
\begin{equation}\label{eqs. 5}
\begin{aligned}
    \sum_j \left[ \left(1 - (1 - \gamma)P_{i}^{A}\right)b_{ji} - \frac{\mu}{\beta^U} I_{ji}\right] \epsilon_j = 0,
\end{aligned}
\end{equation}
where $I_{ji}$ is the element of an identity matrix. Thus, solving Eq. \ref{eqs. 5} reduces to finding the eigenvalues of the matrix whose elements are \(m_{ji} = \left[1 - (1 - \gamma)P_{i}^{A}\right]b_{ji}\), and the minimum \(\beta^{U}\) that satisfies Eq. \ref{eqs. 5} is the critical threshold for the epidemic outbreak. 

Let \(M\) be the matrix with elements \(m_{ji} = \left[1 - (1 - \gamma)P_{i}^{A}\right]b_{ji}\) and the maximum eigenvalue of \(M\) is \(\lambda_{max}(M)\), then the critical threshold \(\beta_{c}^{U}\) is given by:
\begin{equation}\label{eqs. 7}
    \beta_{c}^{U} = \frac{\mu}{\lambda_{max}(M)}.
\end{equation}

According to Eq. \ref{eqs. 7}, we conclude that when information propagation is not considered, the outbreak threshold depends solely on the recovery rate \( \mu \), \( \gamma \), and the structure of the physical layer network. However, when information propagation is taken into account, the outbreak threshold also depends on the proportion of aware nodes \( P_{i}^{A} \).

Theoretically, as we gather information about network structure, infection rates, and other parameters through methods such as sampling surveys, we can construct the corresponding propagation model for theoretical analysis. when calculating the complexity of practical implementation, we need a function to return the neighboring nodes of the node \(i\) in \(r_{i}^{(1)}\), \(r_{i}^{(2)}\), and \(r_{i}^{(3)}\), with the time complexity of \(O(k_i)\), where \(k_i\) is the number of neighbors of the node \(i\). During every state transition process of any given node, the time complexity for updating in the AI, US, and AS states is \(O(1)\), \(O(k_i)\), and \(O(k_i)\), respectively. Therefore, the max time complexity for the update process of all nodes is \(O(n \times \sum_{i=1}^{n} k_i)\), where \(n\) is the number of nodes, and \(k_i\) is the degree of node \(i\). Assuming that the whole state transition process reaches a steady state after at most \( \text{Max\_iteration} \) state transitions, the time complexity one simulation is $O(\text{Max\_iteration} \times (n + \sum_{i=1}^{n} k_i))$. Supposed the final results are getting through average over $s$ simulation times, then the max time complexity for the model becomes $O(s \times \text{Max\_iteration} \times (n + \sum_{i=1}^{n} k_i))$.

As for the scalability of the model and the applicability in real-world data, the research by Gómez et al. indicates that Markov methods are highly usable for this kind of cyber-physical systems \cite{gomez}. Therefore, regardless of whether the network is a synthetic network or a real-world network, the model we construct can effectively simulate the propagation characteristics

From the construction of the nonlinear cyber-physical networked system to the derivation of the epidemic outbreak threshold based on MMCA, we have theoretically developed a novel multiplex network model. Next, we will further investigate our model through simulations.



\section{Simulation}\label{sec: section 3} 
In this section, we examine the fit between theoretical and simulation results, and based on the simulation experiments, explore the coupled dynamics between information propagation and epidemic spreading.

\subsection{Methods}\label{sec: section 3.1}

\begin{figure}[!t]
    \centering
    \includegraphics[width=\linewidth]{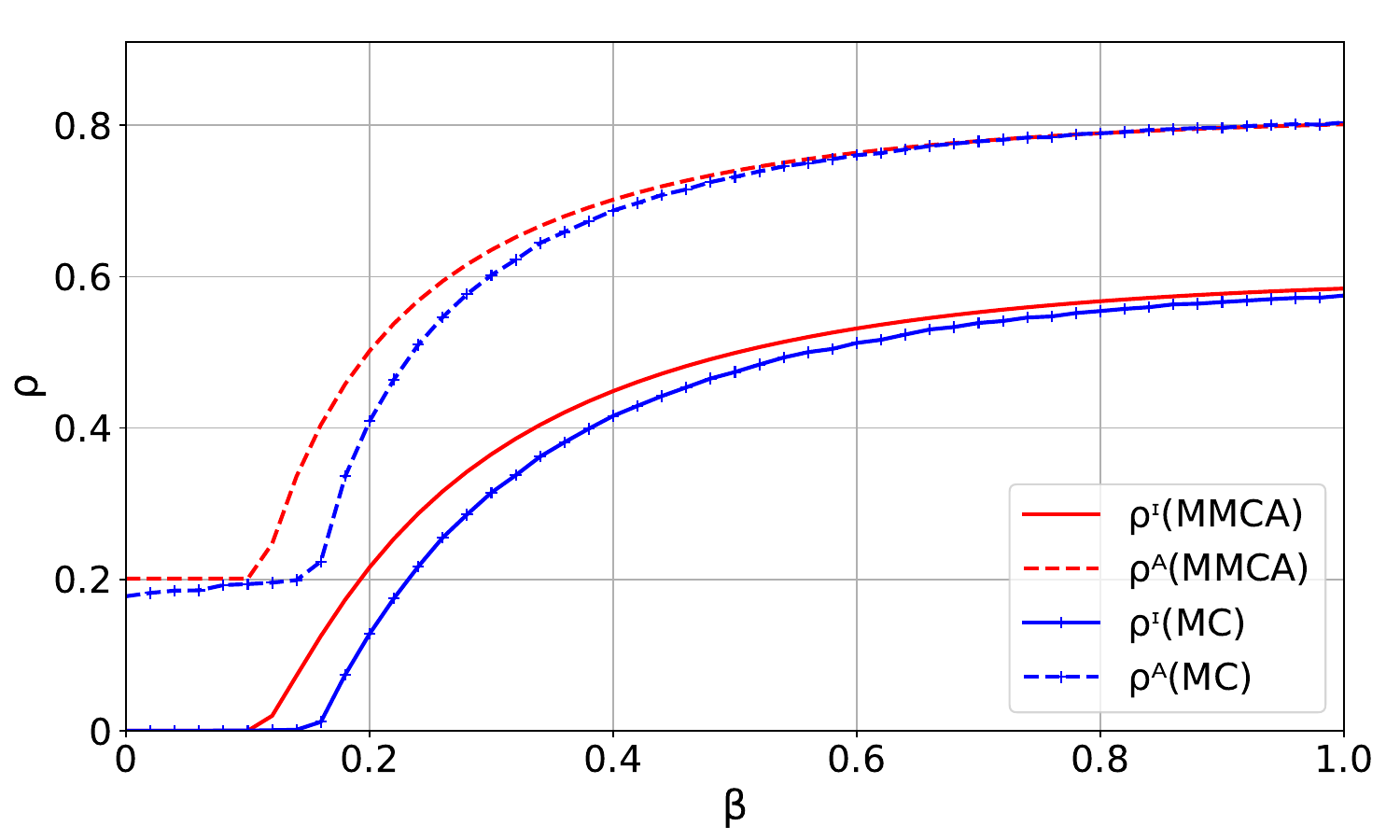}
    \caption{\textbf{Comparison of the stationary density using MMCA and MC methods.} The red solid line and the red dashed line illustrate the infection density \(\rho^{I}\) and the awareness density \(\rho^{A}\) obtained by the MMCA, respectively. The blue solid line with plus markers and the blue dashed line with plus markers display the infection density \(\rho^{I}\) and the awareness density \(\rho^{A}\) obtained by MC simulations, respectively. In the physical layer, the parameters are set as follows: the initial proportion of infected nodes is $1\%$, \(\theta = 0.8\), \(\alpha = 10\), and \(\mu = 0.4\). In the cyber layer, the parameters are set as follows: the average number of 2-simplex each node is part of is $K_{S} = 2$, $\lambda = 0.1$, \(\lambda^{*} = 0.1\), and \(\delta = 0.8\). $\rho^{A} = \frac{\sum_{i}P_{i}^{A}}{N}, \quad \rho^{I} = \frac{\sum_{i}P_{i}^{I}}{N}$. The results are obtained from MC simulations, averaged over 100 iterations, with \(\beta\) varying from 0 to 1 in increments of 0.02.}
    \label{fig. 4}
\end{figure}

In this work, all the simulations are implemented using Python 3.10. 
For the physical layer, the ``Six Degrees of Separation" theory proposed by Stanley Milgram indicates that, on average, only five intermediaries are needed to connect any two strangers in the United States. It highlights two essential characteristics of social networks: short average path length and high clustering coefficient. These features suggest that individuals can connect with others through a few intermediaries while maintaining close-knit relationships with family, friends, or neighbors, forming small groups or communities. The WS small-world network meets these two core characteristics, making it a suitable model for simulating real-world social networks. Hence, we generate a WS network using the \textit{Watts\_strogatz\_graph ()}  function from the \textit{networkx} package and randomly select initially infected nodes using the \textit{random.choice()} function from the \textit{numpy} package. The total number of nodes in the network is $N=1000$ with the initial $1\%$ infected nodes. Each node has an initial number of neighbors $K=4$, and a rewiring probability $P=0.5$. Nodes in the physical layer correspond one-to-one with nodes in the cyber layer. In addition, to simplify the model and focus on investigating the impact of information propagation on cyber-physical systems, we have also neglected potential misjudgment behaviors when individuals attempt to identify the infected ones.

In the current internet era, every individual can act as a self-media, allowing unrestricted dissemination of information beyond physical boundaries. The ER network, characterized by random connections and relatively equal node relationships, effectively simulates information exchange through virtual platforms such as social media and online forums. Therefore, for the cyber layer, we generate an ER random graph using the \textit{erdos\_renyi\_graph ()} function from the \textit{networkx} package and construct 2-simplex using the \textit{combinations()} function from the \textit{itertools} package, with the number of neighbors and the number of 2-simplex determined per node by $p_{1}$ and $p_{2}$ in Sec. \ref{sec: section 2}, respectively, where $k_1=10$ and $k_2=2$. Furthermore, for simplicity, in this work, \( \beta_A \) is set to 0, which means nodes in the A state are immune to the epidemic, namely \( \gamma = 0 \) and \( \beta^U = \beta \).

In the second part of this section, to validate the correctness and completeness of the theory, we simulate the coupled dynamic process of the cyber-physical networked system and compare the results with the theoretical value obtained in Sec. \ref{sec: section 2}. In the rest part of this section, we conduct simulation experiments to study the impact of different types of information propagation on epidemic spreading and discuss their significance for real-world epidemic prevention and control, while we further conduct simulations on power grids, verifying the applicability of our conclusions on real networks.

\subsection{The Comparison between MC Simulation and MMCA}\label{sec: section 3.2}

In this subsection, we simulate the coupled dynamic process between the cyber layer and the physical layer during epidemic spreading using the MC simulation. We plot the steady-state densities of infected and aware nodes under different infection rates $\beta$ to verify our analytical results. The coupled dynamic process in the MC simulation is iterated with parallel updates until the difference between the steady-state densities of two consecutive steps is small enough.

Fig. 4 shows the comparison between the MC simulation and our theoretical prediction of the epidemic threshold $\beta_{c}$. It also can be seen that the agreement between the MMCA and the MC simulation results is reasonably good when calculating the densities of infected and aware nodes, both with an average absolute deviation of around 0.02. Additionally, the results show that the epidemic infection scale varies non-proportionally under different infection rates. This not only indicates that our system is nonlinear, but also highlights that epidemics with higher infection rates lead to larger-scale outbreaks, which can result in more severe social security issues.

\begin{figure*}[!t]
\centering
\subfloat[\(K_{S} = 2\)]{\includegraphics[width=0.325\linewidth]{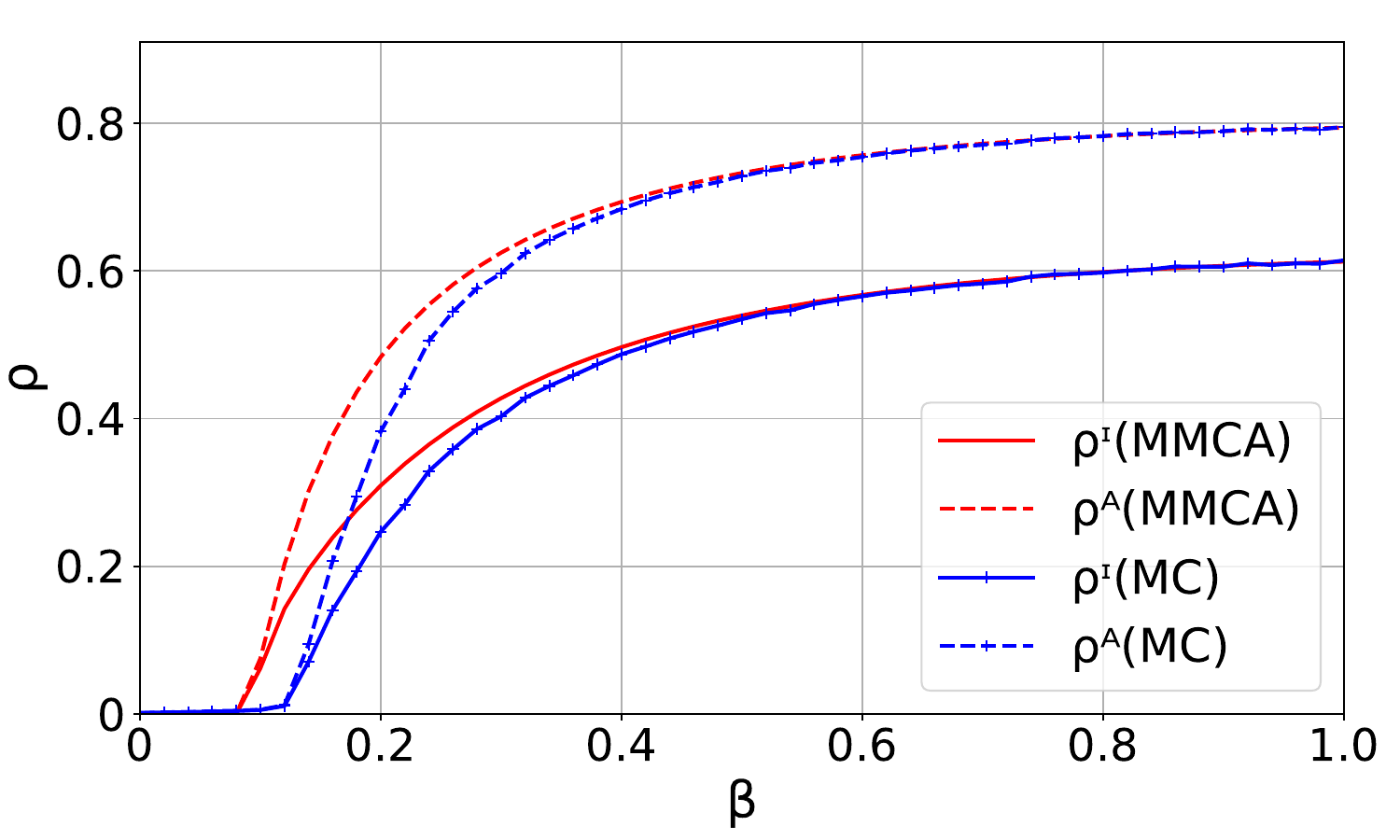}%
\label{fig_first_case}}
\hfil
\subfloat[\(K_{S} = 4\)]{\includegraphics[width=0.325\linewidth]{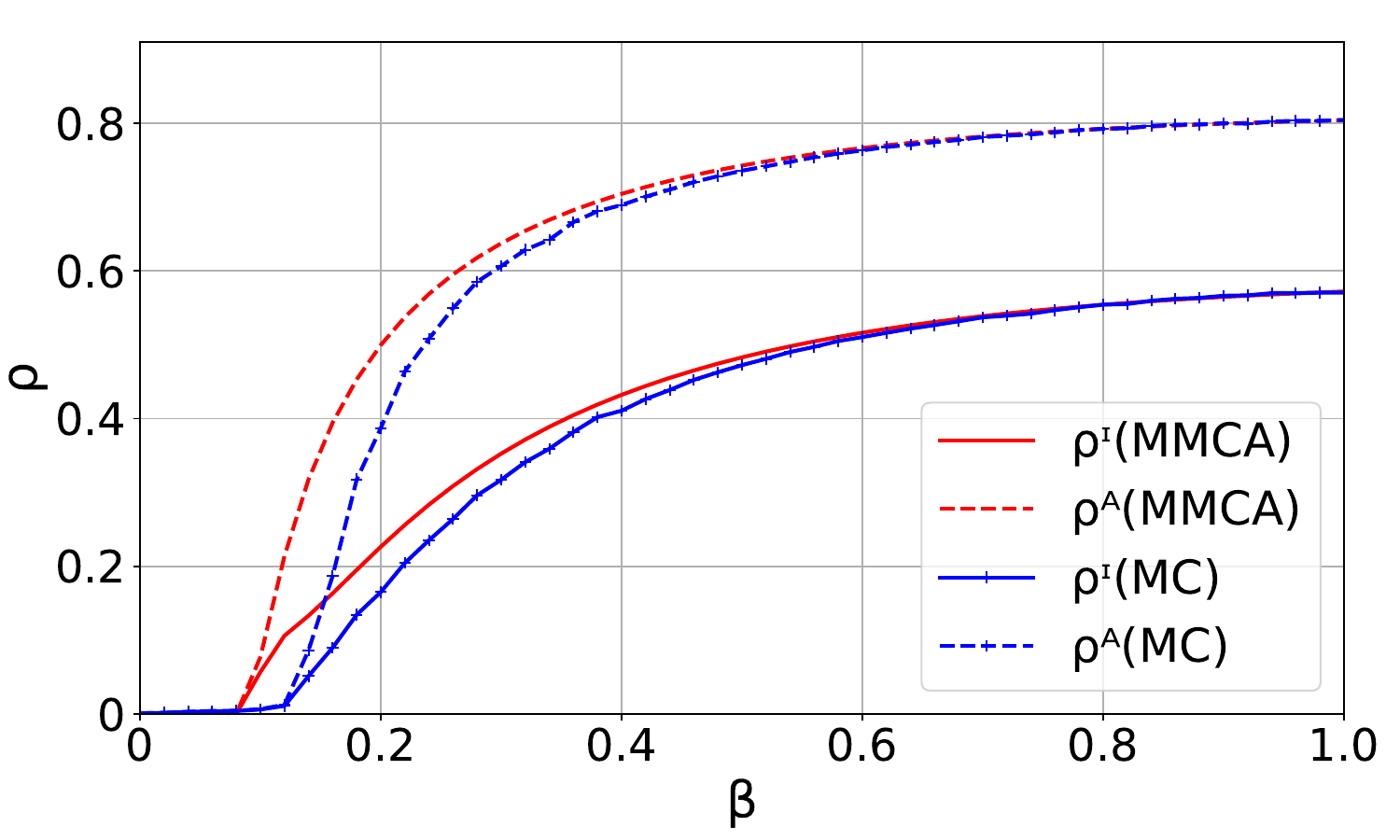}%
\label{fig_second_case}}
\hfil
\subfloat[\(K_{S} = 12\)]{\includegraphics[width=0.325\linewidth]{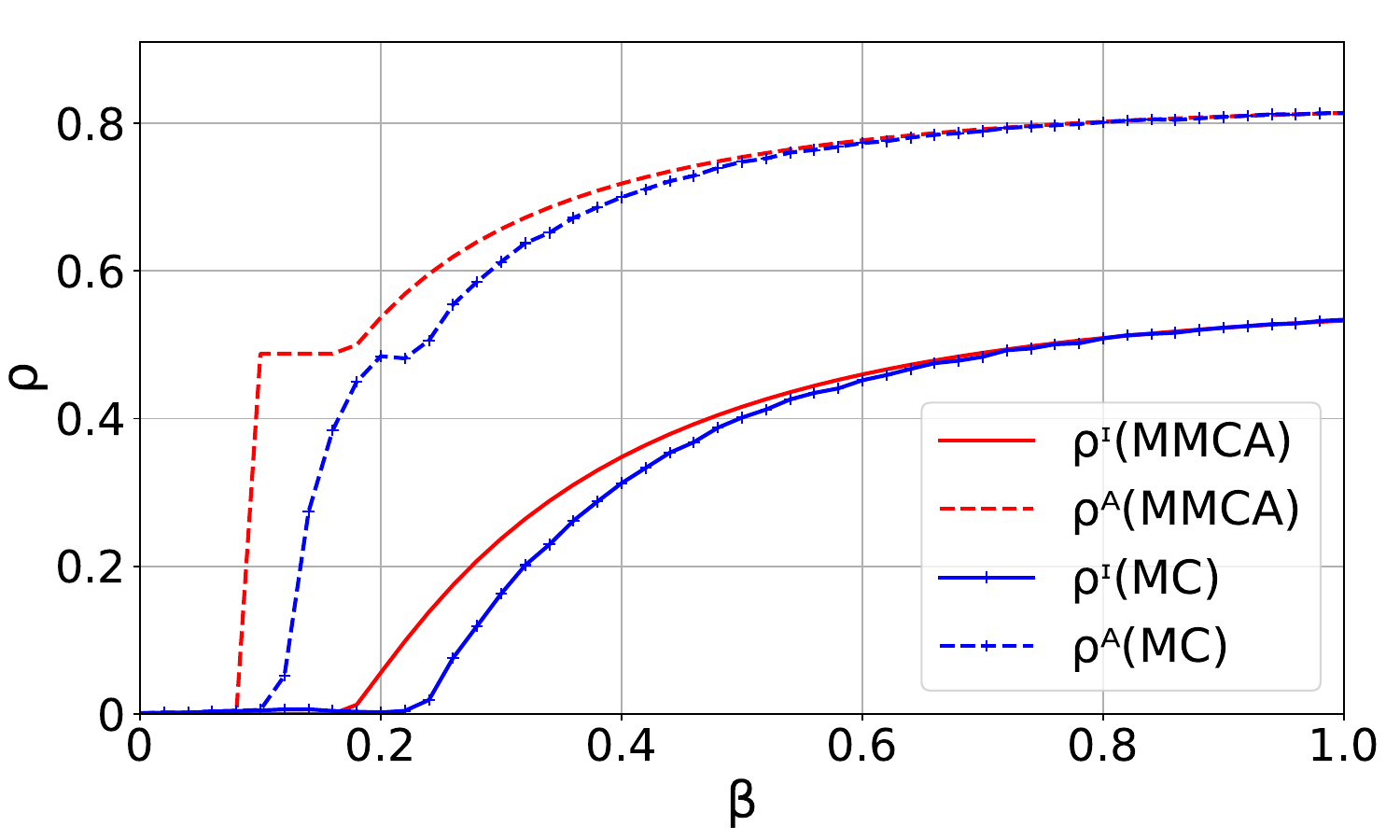}%
\label{fig_third_case}}
\caption{\textbf{Impact of the number of 2-simplex on awareness density and infection density.} The red solid line and the red dashed line illustrate the infection density \(\rho^{I}\) and the awareness density \(\rho^{A}\) obtained by the MMCA, respectively. The blue solid line with plus markers and the blue dashed line with plus markers show the infection density \(\rho^{I}\) and the awareness density \(\rho^{A}\) obtained by MC simulations, respectively. In the physical layer, the initial proportion of infected nodes is 1\%, $r_{i}^{(3)}(t)$ is not considered, and \(\mu = 0.4\). In the cyber layer, $r_{i}^{(1)}(t)$ is not considered, \(\lambda^{*} = 0.5\) and \(\delta = 0.8\). In (a), (b), and (c), the average number of 2-simplex each node is part of is \(K_{S} = 2\), \(K_{S} = 4\), and \(K_{S} = 12\), respectively. The results are obtained from MC simulations, averaged over 100 iterations, with \(\beta\) varying from 0 to 1 in increments of 0.02.}
\label{fig. 5}
\end{figure*}

The slight overestimation of the theoretical values, compared to the simulated ones, can be attributed to the simplified treatment of $P_i^A$ and $r_i^{3}(t)$. In fact, to get the infection threshold theoretically, we need to take the limit of equations in the critical state. As shown in Eq. \ref{eqs. 6} in the original manuscript, as $\beta_{c}^{U}$ near the point of epidemic outbreak, we know that $P_{i}^{AI}$ is very small, but not equal to 0 completely. After taking the limit of \( P_{i}^{A} \), the infection density \( P_{i}^{A} \) will be slightly smaller compared with the actual value. Additionally, the approximation of \( r_{i}^{(3)} \) further reduces \( P_{i}^{A} \) relative to its true value. As a result, the elements of the matrix \(M\) will be slightly larger than their actual values, resulting in the maximum eigenvalue of \(M\) being slightly higher than the actual value, and this causes the theoretical infection threshold derived from Eq. \ref{eqs. 7}, to be lower than the real threshold. All of those above will contribute to the theoretical infection threshold being consistently slightly lower than the actual threshold. This explains why the infection curve obtained by MMCA does not perfectly align with the MC infection curve. 

Furthermore, we should note that when infection rates are below the MMCA theoretical infection threshold, the epidemic will not spread. However, when the initial number of infections is small but the infection rate is clearly above the infection threshold, the epidemic will still break out, though at a relatively slower pace. This is because the transmission process is driven by both the infection spread and recovery processes. The infection rate \(\beta\) represents the rate at which each infected node can infect susceptible nodes per unit time, while the recovery rate \(\mu\) represents the rate at which each infected node recovers to a susceptible state per unit time. Therefore, as long as the overall infection rate in the system is higher than the recovery rate, the infection will spread among neighbors. Furthermore, with a low initial number of infected individuals, infected individuals are able to come into contact with a larger number of susceptible individuals, leading to an epidemic outbreak. After the epidemic spreads, the proportion of susceptible individuals decreases significantly, reducing the contact opportunities between infected and susceptible individuals, thus slowing the epidemic's spread. When the number of newly infected individuals equals the number of individuals recovering to the susceptible state, the infection ratio reaches a steady state. Therefore, the initial number of infected individuals does not directly affect whether an epidemic will occur or its scale, but it can significantly accelerate the speed at which the epidemic breaks out.

\begin{figure*}[!t]
\centering
\subfloat[Awareness density with \boldmath{$\alpha = 10$}]{\includegraphics[width=0.5\linewidth]{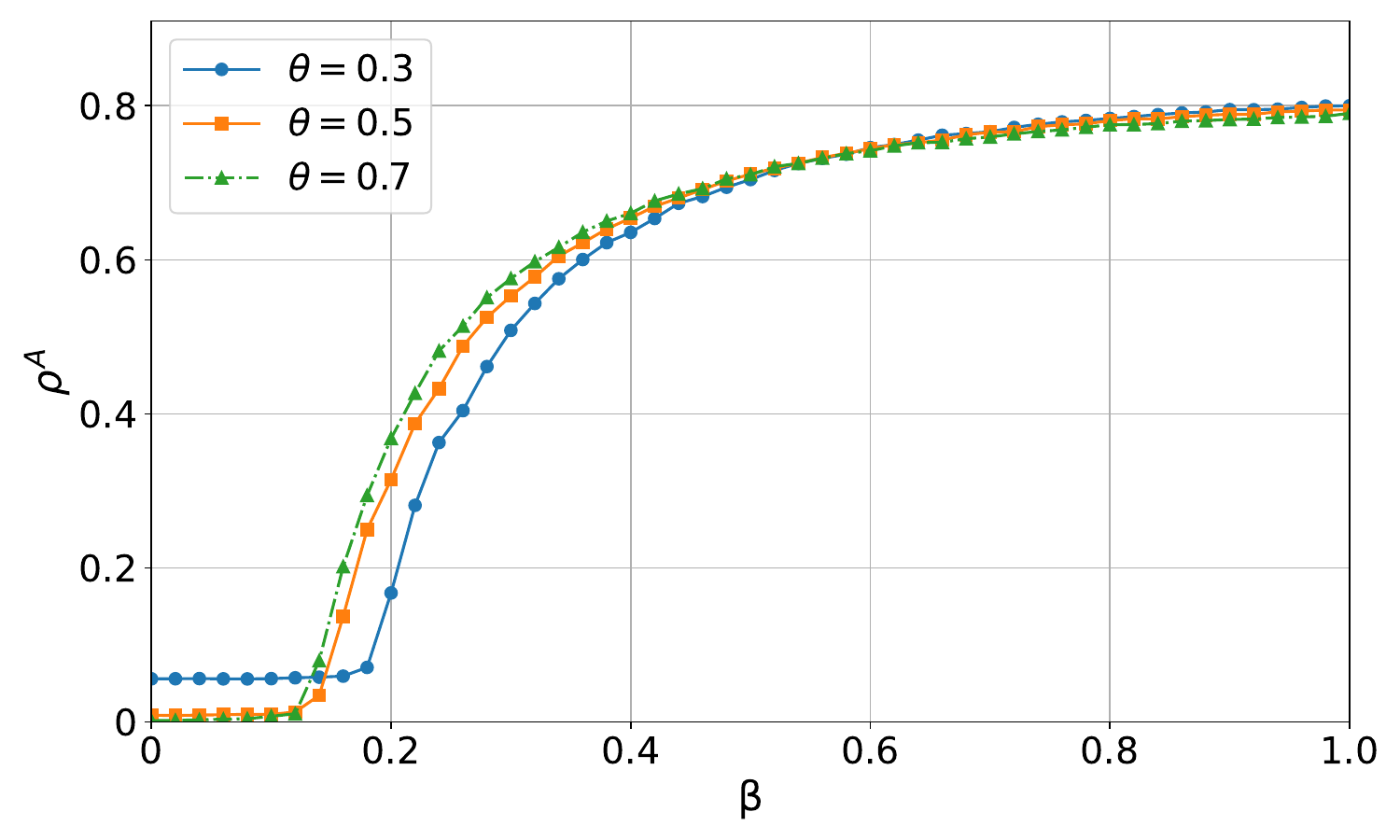}%
\label{fig_first_case}}
\hfil
\subfloat[Infection density with \boldmath{$\alpha = 10$}]{\includegraphics[width=0.5\linewidth]{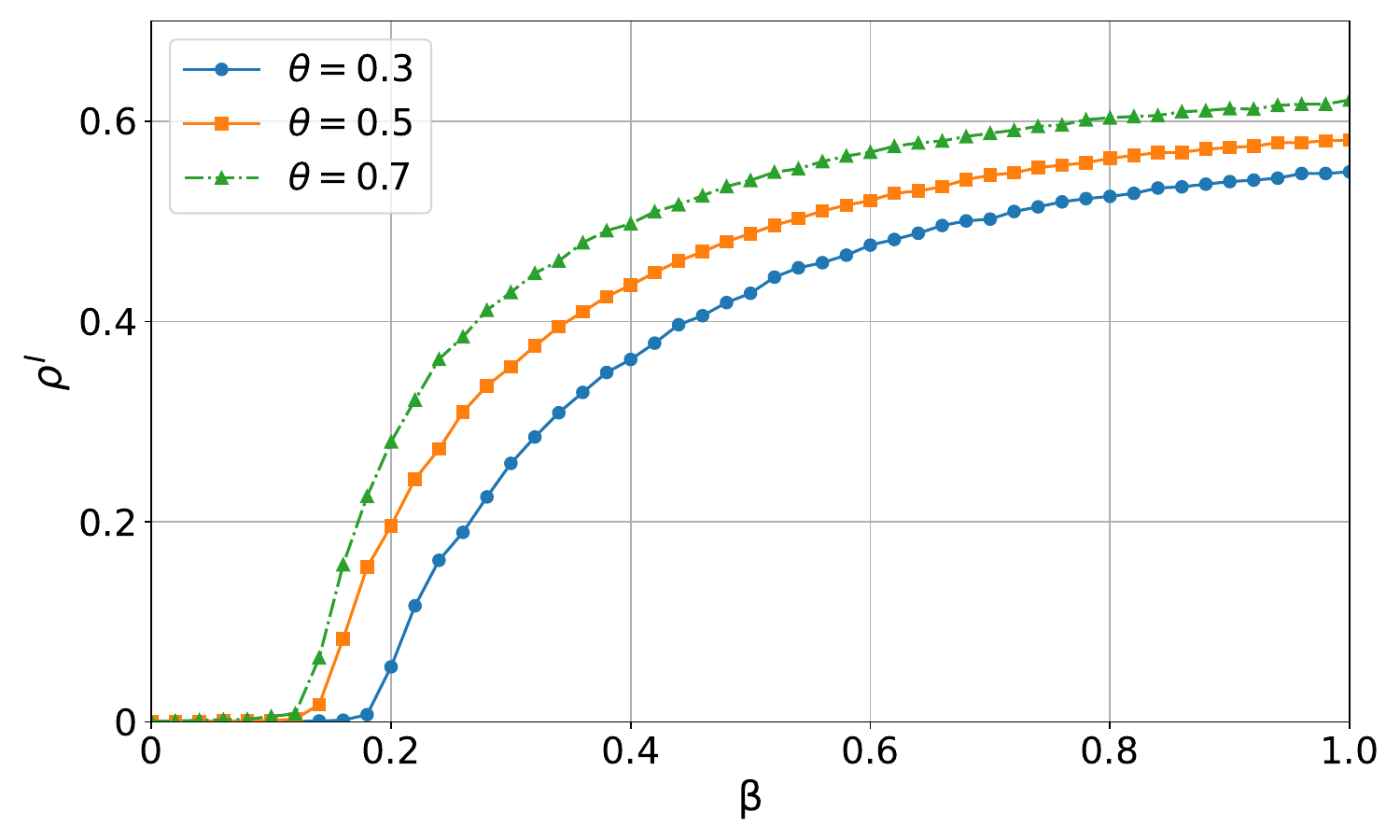}%
\label{fig_second_case}}
\hfil
\subfloat[Awareness density with \boldmath{$\alpha = 15$}]{\includegraphics[width=0.5\linewidth]{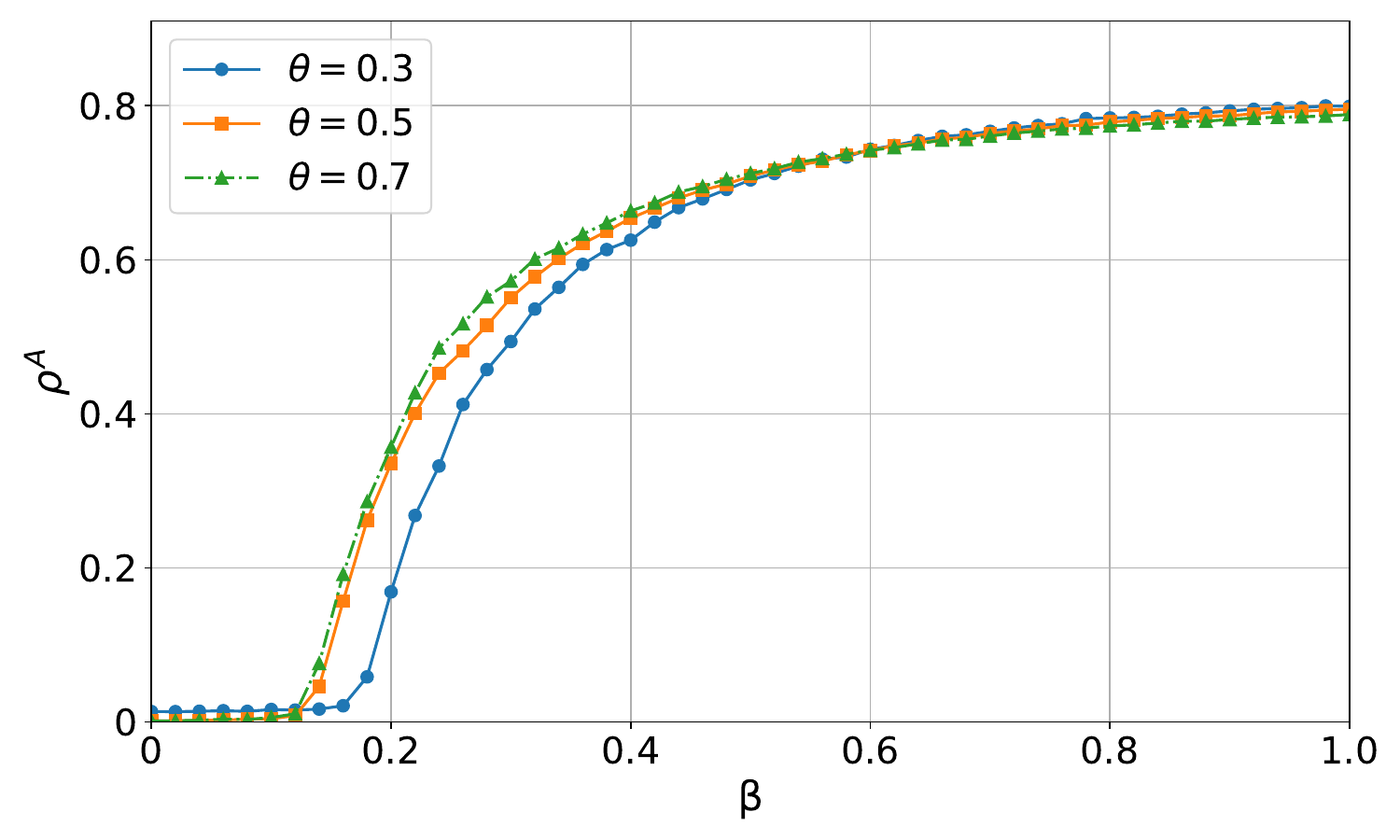}%
\label{fig_t_case}}
\hfil
\subfloat[Infection density with \boldmath{$\alpha = 15$}]{\includegraphics[width=0.5\linewidth]{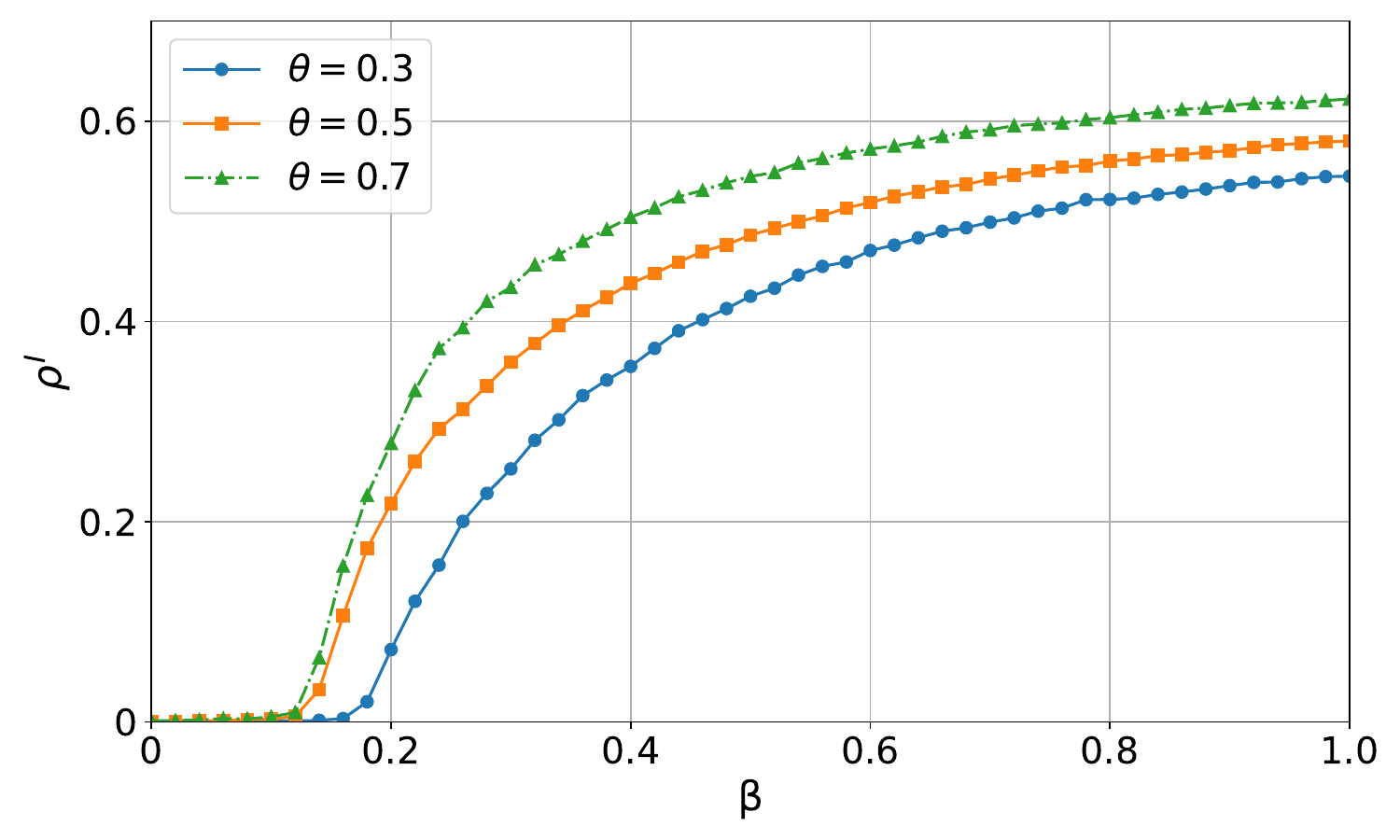}%
\label{fig_f_case}}
\caption{\textbf{Impact of physical-layer information on awareness density and infection Density.} The blue dotted solid line, yellow square dotted solid line, and green triangular dash-dotted line display the awareness density $\rho^{A}$ or the infection density \(\rho^{I}\) obtained by the MC simulations for \(\theta = 0.3\), \(\theta = 0.5\), and \(\theta = 0.7\), respectively. In (a) and (b), the curves denote the awareness density \(\rho^{A}\) and infection density \(\rho^{I}\) for response strength \(\alpha = 10\), while in (c) and (d), the curves denote the awareness density \(\rho^{A}\) and infection density \(\rho^{I}\) for response strength \(\alpha = 15\). In the physical layer, the initial proportion of infected nodes is 1\%, and \(\mu = 0.4\). In the cyber layer, $r_{i}^{(1)}(t)$ and  $r_{i}^2(t)$ are not considered, and \(\delta = 0.8\). The results are obtained from MC simulations, averaged over 100 iterations, with \(\beta\) varying from 0 to 1 in increments of 0.02.}
\label{fig. 6}
\end{figure*}

The simulations demonstrate that the MMCA effectively predicts the infection rate in the network and the proportion of the population aware of the epidemic. To gain a more comprehensive understanding of the coupled dynamics between information propagation and epidemic spreading, we will explore the impact of different types of information propagation on epidemic spreading in the following section.

\begin{figure*}[!t]
\centering
\subfloat[Awareness density with pairwise interaction]{\includegraphics[width=0.487\linewidth]{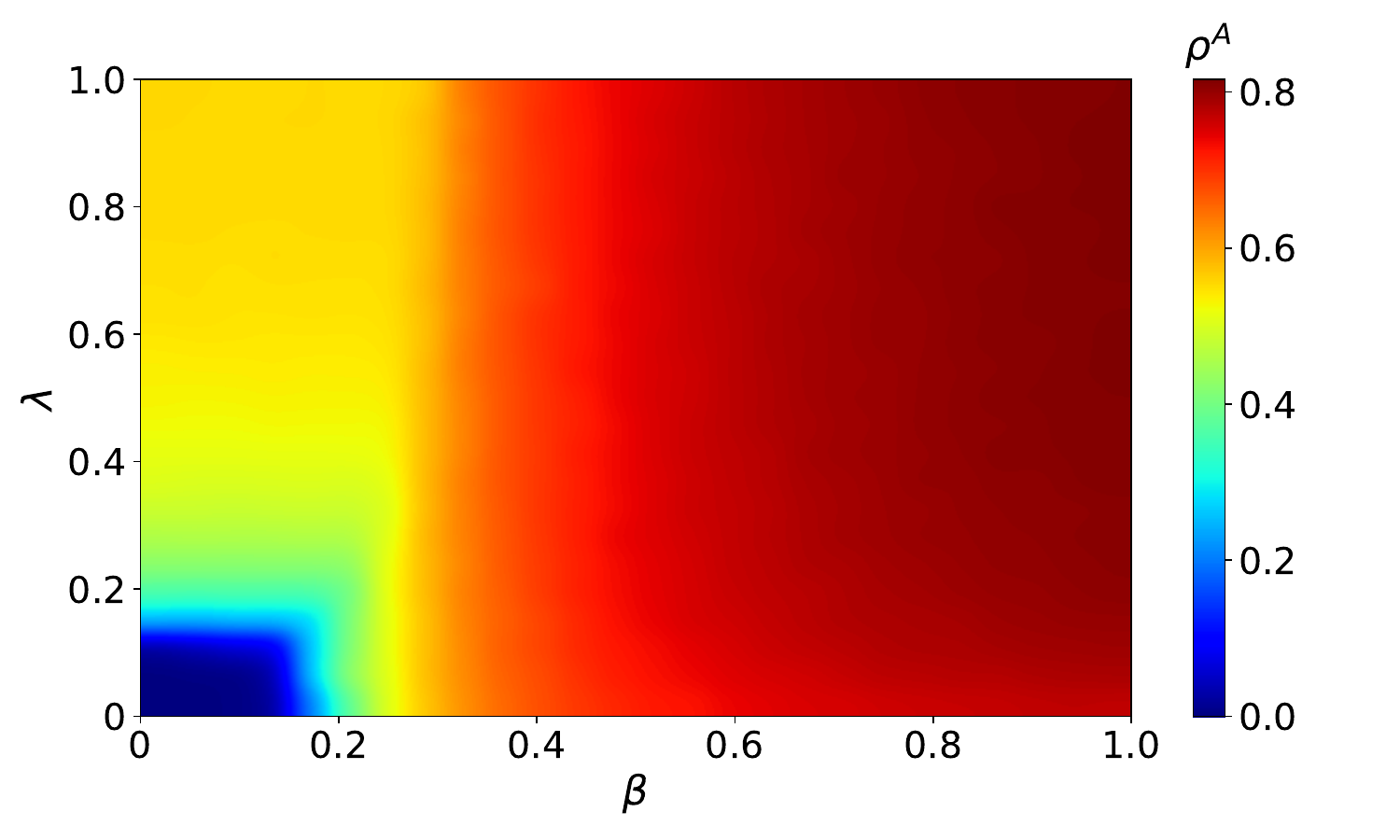}%
\label{fig_first_case}}
\hfil
\subfloat[Infection density with pairwise interaction]{\includegraphics[width=0.487\linewidth]{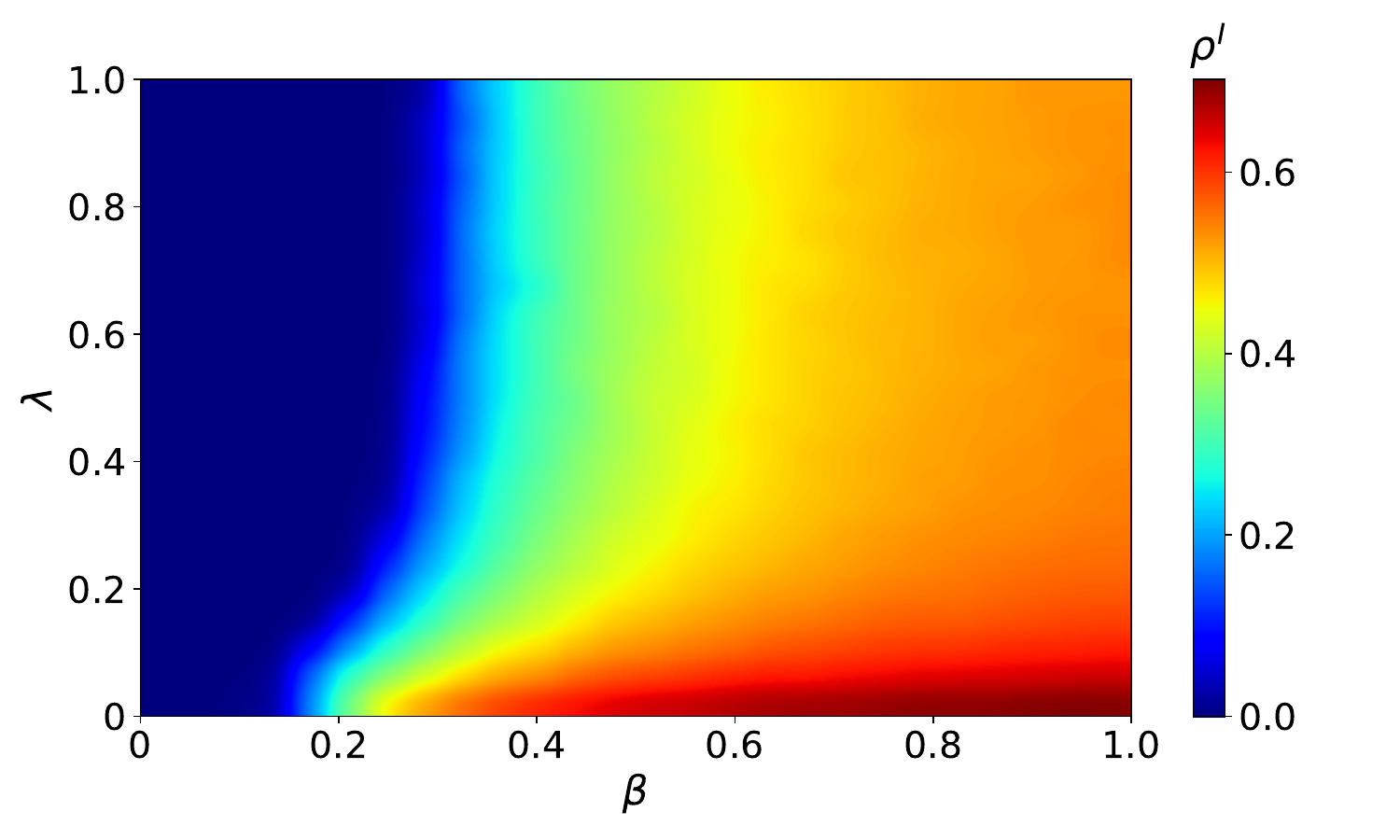}%
\label{fig_second_case}}
\hfil
\subfloat[Awareness density with 2-simplex information]{\includegraphics[width=0.487\linewidth]{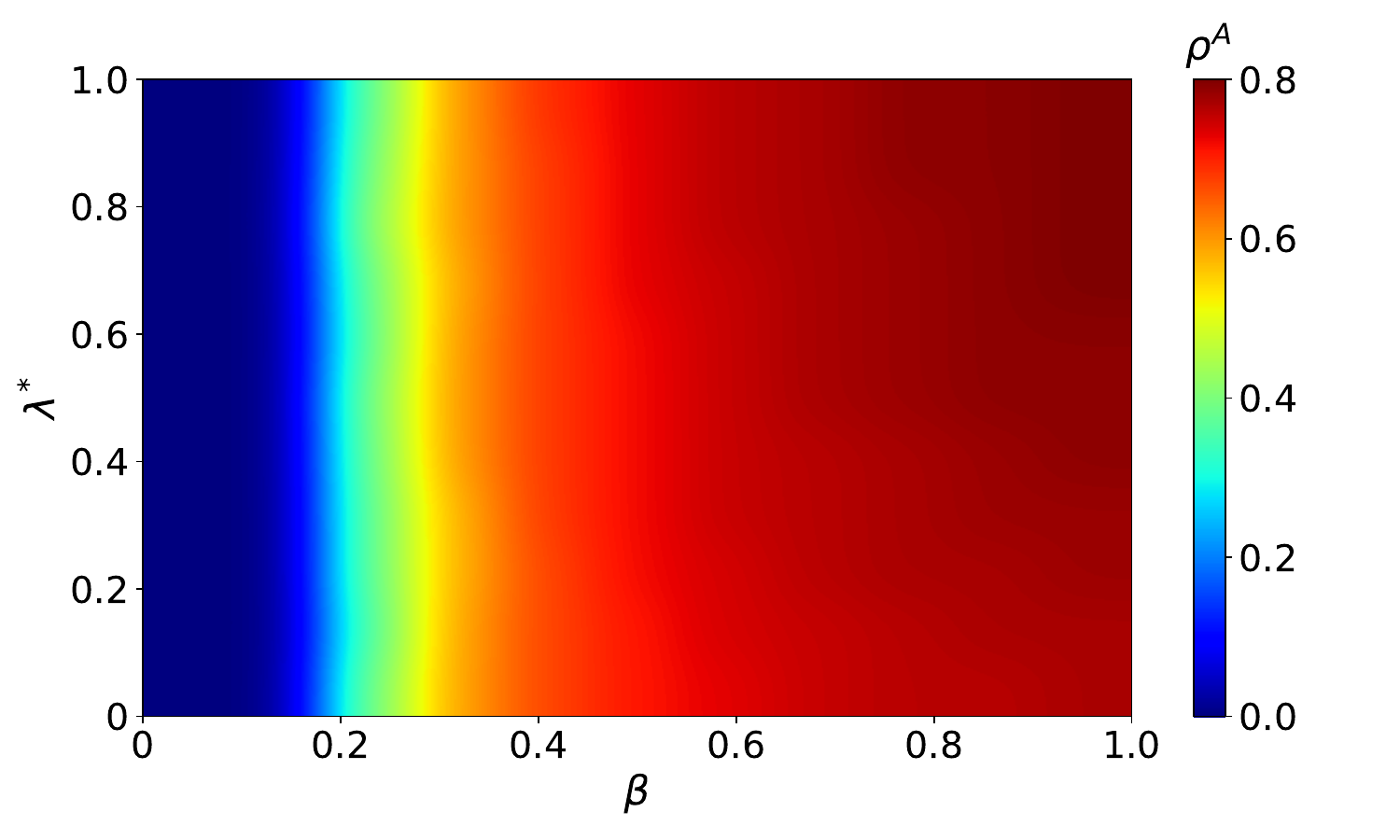}%
\label{fig_t_case}}
\hfil
\subfloat[Infection density with 2-simplex information]{\includegraphics[width=0.487\linewidth]{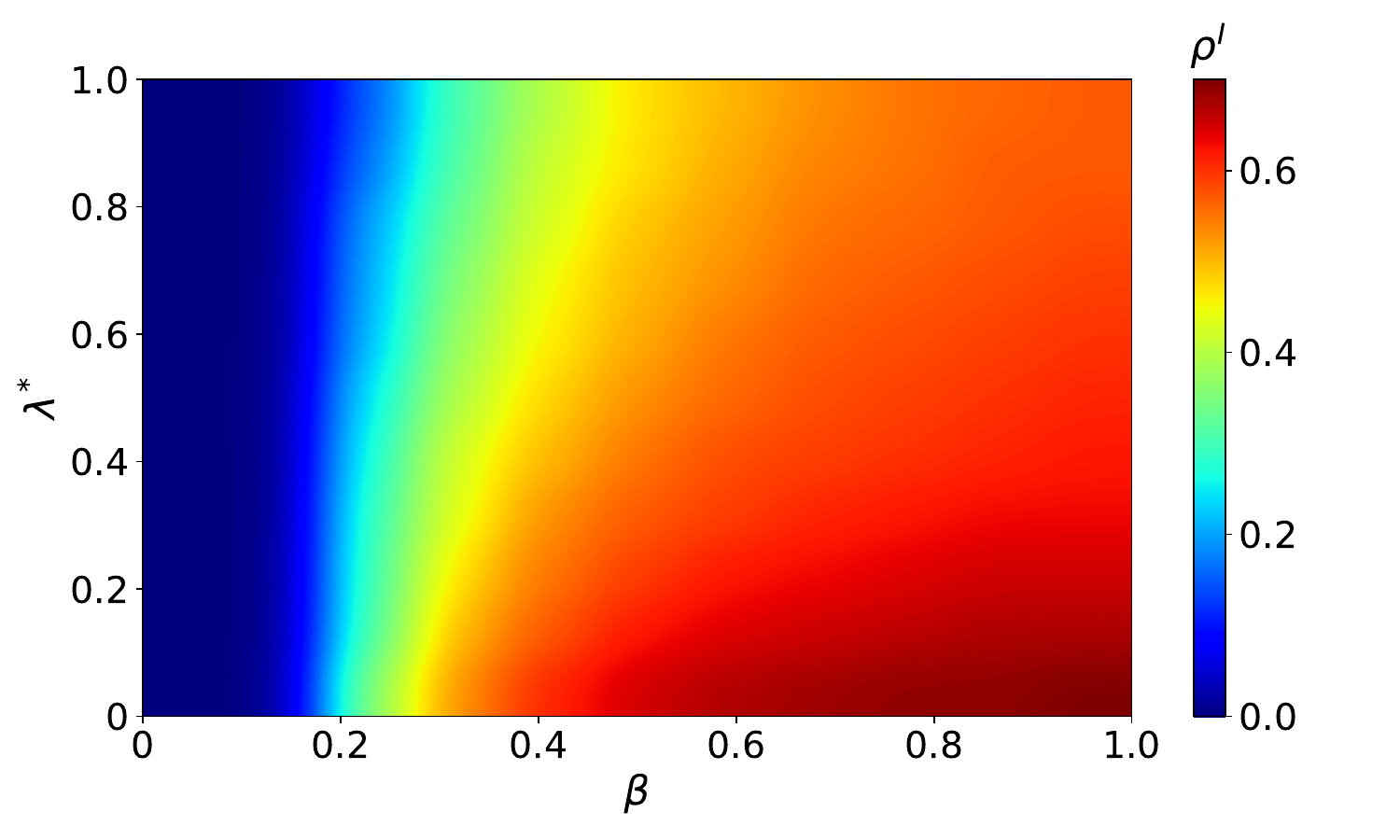}%
\label{fig_f_case}}
\hfil
\subfloat[Awareness density with physical-layer information]{\includegraphics[width=0.487\linewidth]{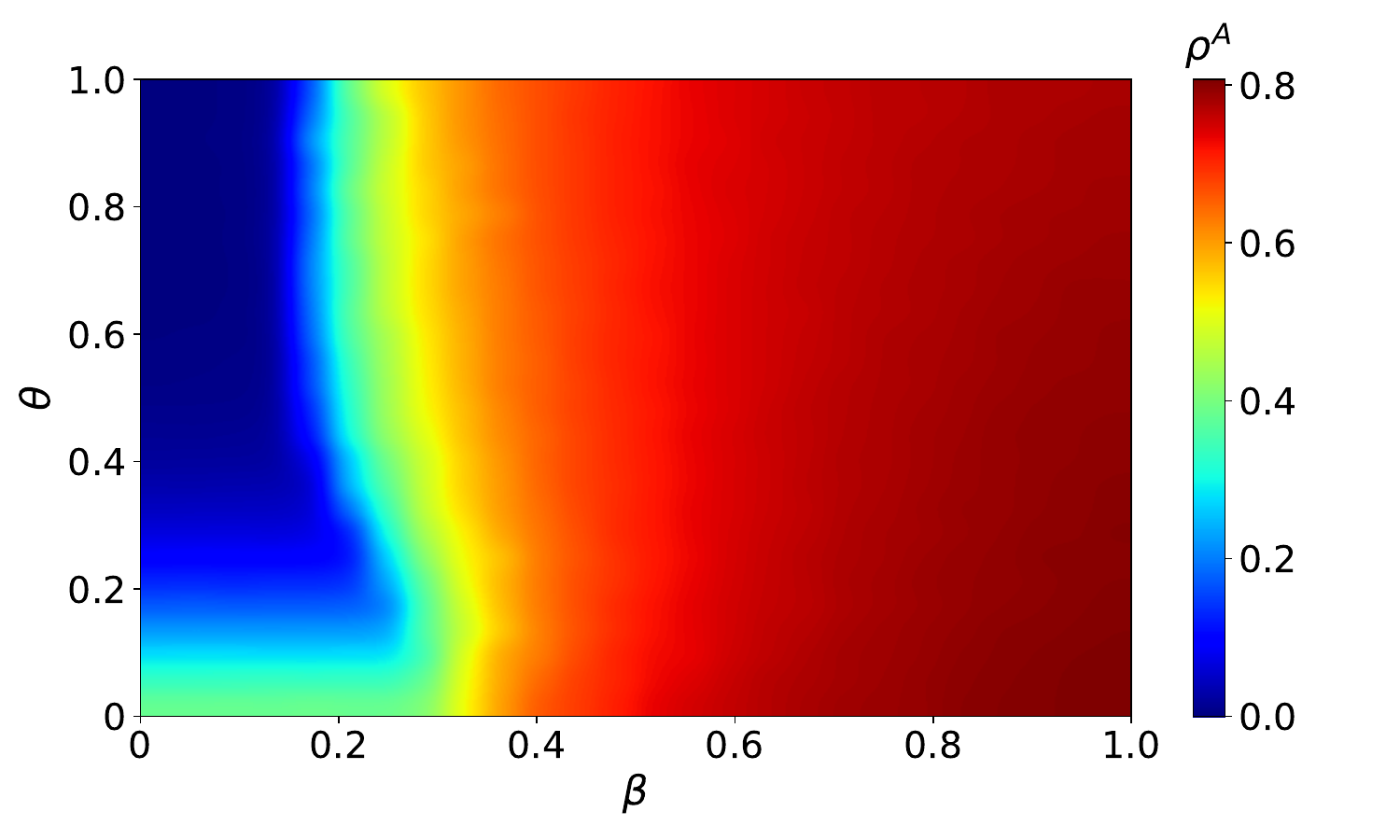}%
\label{fig_f_case}}
\hfil
\subfloat[Infection density with physical-layer information]{\includegraphics[width=0.487\linewidth]{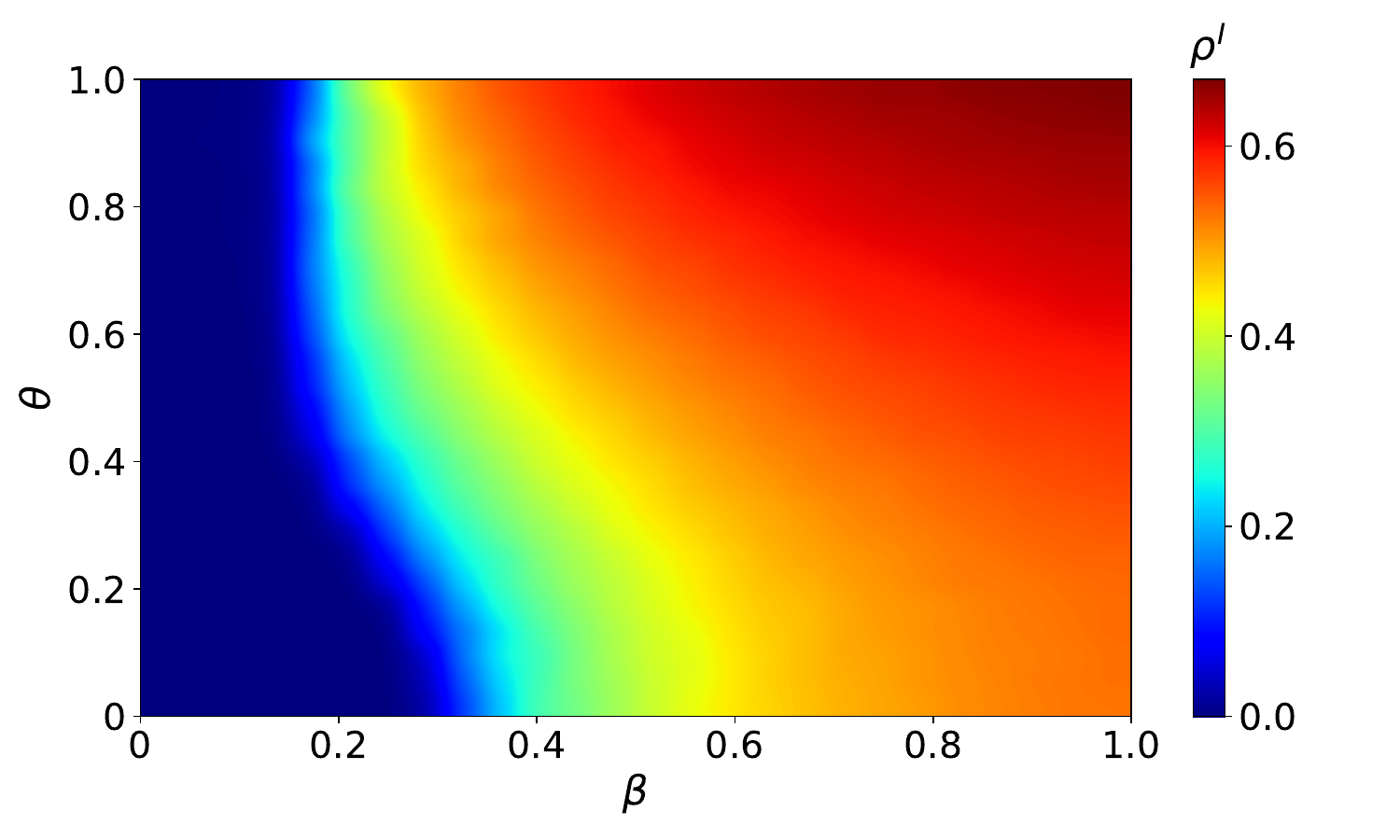}%
\label{fig_f_case}}
\caption{\textbf{Heatmaps of awareness density and infection density under different information propagation methods.} (a) and (b) show the awareness density and infection density when only pairwise interaction information is considered. (c) and (d) illustrate the awareness density and infection density when only 2-simplex information is considered. (e) and (f) display the awareness density and infection density when only physical-layer information propagation is considered. In (a) and (b), in the physical layer, the initial proportion of infected nodes is 1\%, \(r_{i}^{3}(t)\) is not considered, and \(\mu = 0.4\). In the cyber layer, \(r_{i}^{2}(t)\) is not considered, and \(\delta = 0.8\); In (c) and (d), in the physical layer, the initial proportion of infected nodes is 1\%, \(r_{i}^{3}(t)\) is not considered, and \(\mu = 0.4\). In the cyber layer, \(r_{i}^{1}(t)\) is not considered, $K_{S}=2$ and \(\delta = 0.8\); In (e) and (f), in the physical layer, the initial proportion of infected nodes is 1\%, \(\alpha = 10\), and \(\mu = 0.4\). In the cyber layer, \(r_{i}^{1}(t)\) and \(r_{i}^{2}(t)\) are not considered, and \(\delta = 0.8\). All the results are obtained from MC simulations, averaged over 100 iterations. The color of each panel consists of the density in a 50 × 50 grid.}
\label{fig. 7}
\end{figure*}

\begin{figure*}[!t]
\centering
\subfloat[Awareness density]{\includegraphics[width=0.485\linewidth]{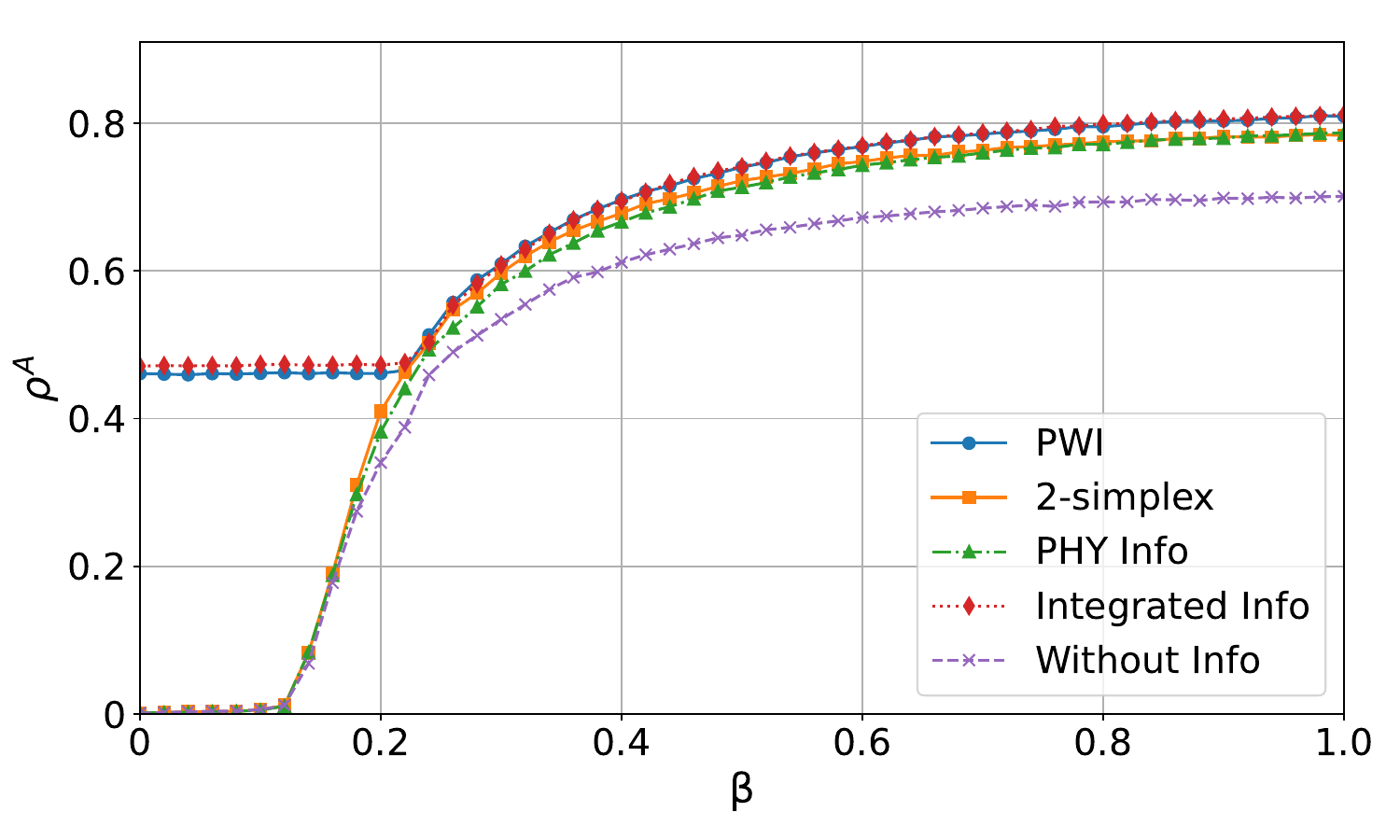}%
\label{fig_first_case}}
\hfil
\subfloat[Infection density]{\includegraphics[width=0.485\linewidth]{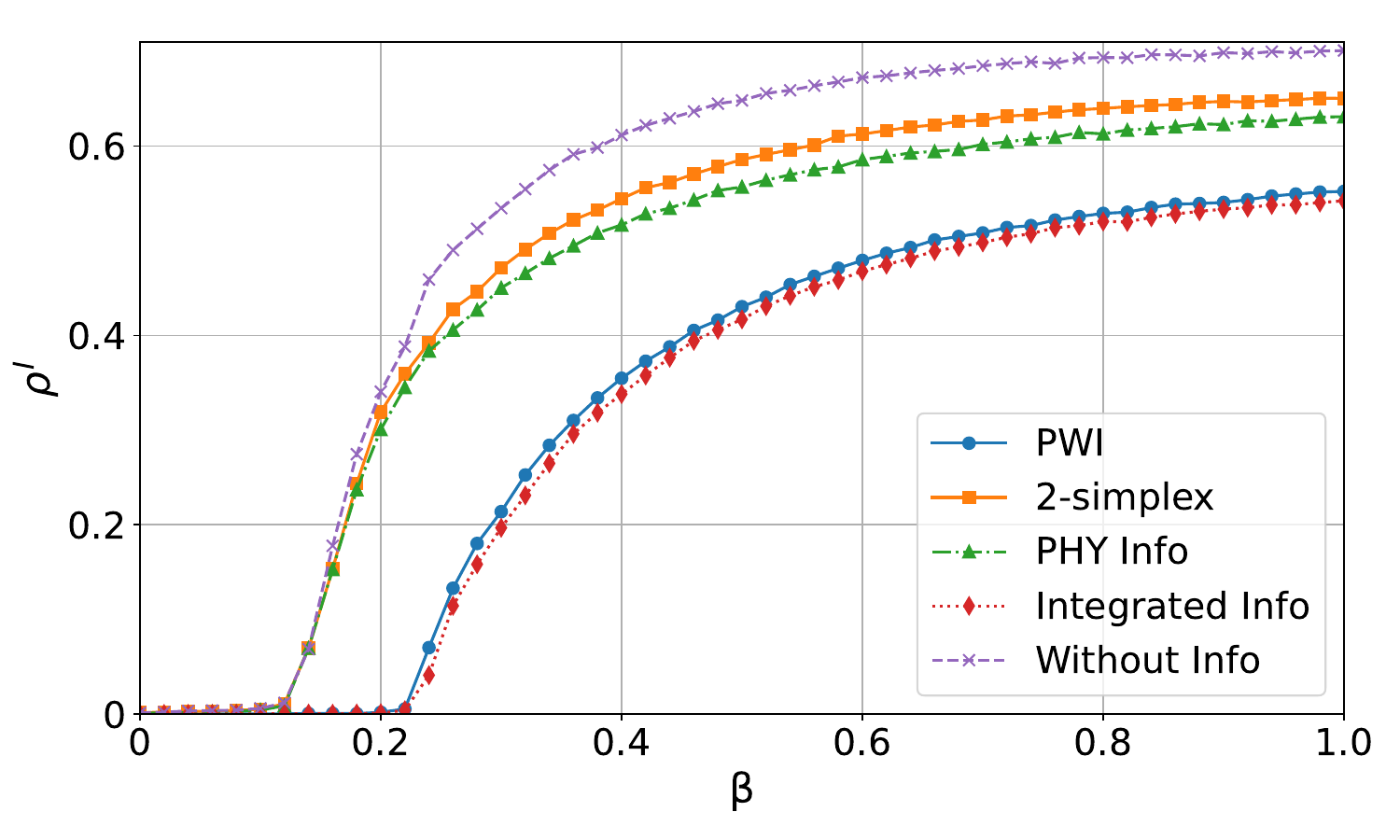}%
\label{fig_second_case}}
\caption{\textbf{Impact of various information on awareness density and infection density.} The blue solid line with circle markers, orange solid line with square markers, green dash-dot line with triangle markers, red dotted line with diamond markers, and purple dashed line with X markers denote infection density $\rho^{I}$ and the awareness density $\rho^{A}$ under the condition of pairwise information (PWI), 2-simplex information, physical information (PHY Info), integrated information (Integrated Info), and without information (Without Info), respectively. As for the condition of Integrated Info, in the physical layer, the initial proportion of infected nodes is 1\%, $\theta = 0.76$, $\alpha = 10$, and $\mu = 0.4$. In the cyber layer, $\lambda = 0.24$, $\lambda^{*} = 0.24$, and $\delta = 0.8$. In the circumstance of PWI, 2-simplex, and PHY Info, only the information represented by $r_i^{(1)}$, $r_i^{(2)}$, and $r_i^{(3)}$, respectively, is considered. In the condition of Without Info, there is no propagation of information during the epidemic. The results are obtained from MC simulations, averaged over 100 iterations, with $\beta$ varying from 0 to 1 in increments of 0.02.}
\label{fig. 8}
\end{figure*}

To explore the impact of different types of information propagation on the epidemic, we present the roles of 2-simplex and physical-layer information in epidemic spreading in Figs. \ref{fig. 5} and \ref{fig. 6}, respectively, and compare the effects of pairwise interaction, 2-simplex, and physical-layer information on suppressing epidemic spreading and promoting information propagation in Figs. \ref{fig. 7} and \ref{fig. 8}.

\subsection{The Impact of Information Propagation on Awareness Density and Infection Density}\label{sec: section 3.3}
In Fig. \ref{fig. 5}, we show the curves of infected and aware node densities as a function of the infection rate $\beta$ when each node is located in 2, 4, and 12 2-simplices on average. The average absolute deviation between the MC simulation and the MMCA is around 0.024.

Comparing Figs. \ref{fig. 5}(a) and (b), we observe that as the average number of 2-simplex per node increases from 2 to 4, the infection threshold $\beta_{c}$ also increases with the number of 2-simplex. When the infection rate $\beta$ is below the infection threshold $\beta_{c}$, both the infection density and awareness density are zero. However, when the infection rate $\beta$ exceeds the infection threshold $\beta_{c}$, the steady-state infection density decreases while the steady-state awareness density increases, the phenomenon is more evident in Fig. \ref{fig. 5}(c). Although Iacopini et al. have explored the role of 2-simplex in suppressing epidemic spread through peer pressure and reinforcement effects with significant results, they did not pay attention to the impact and effectiveness of simplex density in promoting information diffusion \cite{simplicial}, \cite{fan}, and in Fig. \ref{fig. 5}(c), we notice an interesting phenomenon: when the infection rate $\beta$ is below the infection threshold $\beta_{c}$, the steady-state distribution of awareness density exhibits a biphasic region. The first segment shows a steady awareness density of 0, while the second segment shows a steady awareness density of approximately 0.5. It occurs because, in the absence of pairwise interactions and physical-layer information, although 2-simplex information propagation is relatively stringent, the average number of 2-simplex surrounding an individual in Fig. \ref{fig. 5}(c) significantly increases compared to Fig. \ref{fig. 5}(b). Under low infection rates $\beta$, the probability of information diffusion remains limited due to the stringent 2-simplex information propagation conditions; however, when the infection rate $\beta$ is higher (still below the infection threshold $\beta_{c}$), the corresponding increase in awareness density in the cyber layer greatly enhances the probability of information propagation through 2-simplex, and at this point, 2-simplex information propagation functions similarly to the pairwise interaction, reaching the diffusion threshold for the pairwise interaction information propagation, and the information diffuses. When the infection rate $\beta$ exceeds the infection threshold $\beta_{c}$, the epidemic breaks out, leading to an increase in infection density, which correspondingly raises the awareness density. 

From Figs. \ref{fig. 5}(a) to (c), we observe that 2-simplex information propagation increases the infection threshold by approximately 0.1, raises the steady-state awareness density by about 3\% on average, and reduces the steady-state infection density by roughly 12\% on average. Next, we will explore the role and effects of physical-layer information in epidemic spreading.

A low individual epidemic vigilance threshold $\theta$ implies that even with a low density of infected nodes, individuals are likely to take protective measures to avoid infection. Observing Figs. \ref{fig. 6}(a) and (b), we find that a higher epidemic vigilance threshold $\theta$ corresponds to a lower infection threshold $\beta_{c}$ and a higher steady-state infection density. However, the individual epidemic vigilance threshold $\theta$ has little impact on the steady-state awareness density, especially when the infection rate $\beta$ is high. This is because the physical-layer information provides a source for information propagation in the cyber layer, and as the infection rate $\beta$ increases, more information sources become available, leading to saturation in information propagation within the cyber layer. Comparing Figs. \ref{fig. 6}(a) and (c), we observe that the overall impact of response strength $\alpha$ on epidemic information propagation is limited, as it has a weak effect on decreasing the infection density and increasing the awareness density of the population and the infection threshold. A similar conclusion can be drawn from comparing Figs. \ref{fig. 6}(b) and (d), where the response strength $\alpha$ has a minimal effect on suppressing the spreading of the epidemic.

In Fig. \ref{fig. 6}, as $\theta$ increases from 0.3 to 0.7, we observe that greater sensitivity to the epidemic—i.e., lowering the individual epidemic vigilance threshold $\theta$—raises the infection threshold $\beta_c$ by approximately 0.5 and decreases the steady-state infection density by about 10\%. Additionally, individual response strength $\alpha$ plays a role in epidemic spreading, such as decreasing the infection density and increasing the awareness density of the population and the threshold of infection, but its impact is far less significant than that of the individual epidemic vigilance threshold $\theta$. In summary, we have discussed the impact of 2-simplex and physical-layer information propagation on epidemic spreading, then we will examine the combined effects of pairwise interaction, 2-simplex, and physical-layer information.

\subsection{A Comparative Study on the Impact of Different Types of Information on Epidemic Spreading}\label{sec: section 3.4}
We have explored the impact of different types of information on epidemic spreading. In this subsection, we compare and analyze the role of this information in suppressing epidemic spreading. 

\begin{figure}[!t]
    \centering
    \includegraphics[width=\linewidth]{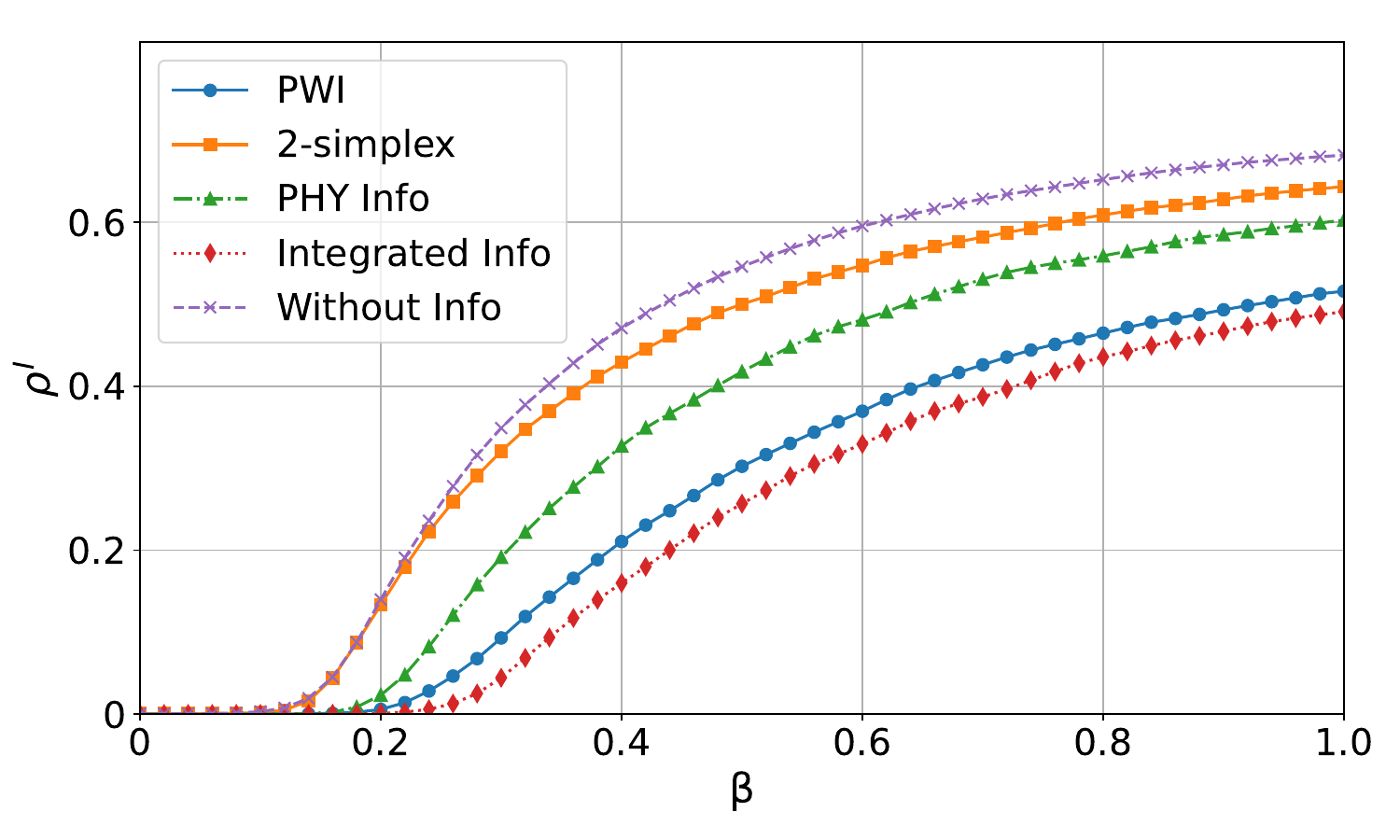}  
    \caption{\textbf{Impact of various information on the affected density in the power grid.} The blue solid line with circle markers, orange solid line with square markers, green dash-dot line with triangle markers, red dotted line with diamond markers, and purple dashed line with X markers denote infection density $\rho^{I}$ under the condition of pairwise information (PWI), 2-simplex information, physical information (PHY Info), integrated information (Integrated Info), and without information (Without Info), respectively. As for the condition of Integrated Info, in the physical layer, the initial proportion of infected nodes is 1\%, $\theta = 0.1$, $\alpha = 10$, and $\mu = 0.4$. In the cyber layer, $\lambda = 0.24$, $\lambda^{*} = 0.24$, and $\delta = 0.8$. In the circumstance of PWI, 2-simplex, and PHY Info, only the information represented by $r_i^{(1)}$, $r_i^{(2)}$, and $r_i^{(3)}$, respectively, is considered. In the condition of Without Info, there is no propagation of information during the epidemic. The results are obtained from MC simulations, averaged over 100 iterations, with $\beta$ varying from 0 to 1 in increments of 0.02.}
    \label{fig. 9}
\end{figure}

In Fig. \ref{fig. 7}, we present heatmaps of awareness density and infection density under different infection rates $\beta$ as a function of $\lambda$, $\lambda^{*}$, and $\theta$, which are conducted without considering the influence of the other two types of information. In Fig. \ref{fig. 7}(a), we observe a rectangular region in the lower left corner of the heatmap where the awareness density is 0, since when the pairwise interaction information propagation rate in the cyber layer is below the information outbreak threshold, information cannot propagation. However, when the infection rate $\beta$ exceeds the infection threshold, the outbreak of the epidemic provides information sources for the cyber layer, and information spreads, forming the rectangular region in Fig. \ref{fig. 7}(a). Comparing Figs. \ref{fig. 7}(a), (c), and (e), we find that the higher the infection rate $\beta$, the greater the awareness density. The larger the values of $\lambda$ and $\lambda^{*}$ or the lower the value of $\theta$, the higher the awareness density. The differences in the effects of $\lambda$, $\lambda^{*}$, and $\theta$ on awareness density become less significant when $\beta > 0.2$. In Figs. \ref{fig. 7}(a), (c), and (e), the average awareness densities are around 0.678, 0.584, and 0.573, respectively, which indicates that both three ways of information propagation are effective in promoting the awareness of people. In Fig. \ref{fig. 7}(b), when the infection rate $\beta$ is below the infection threshold (approximately 0.2), we observe that the epidemic cannot spread. When the infection rate $\beta$ exceeds the infection threshold, and as $\beta$ increases while $\lambda$ decreases, the infection density increases. Similar conclusions can be drawn from Figs. \ref{fig. 7}(d) and (f). Comparing Figs. \ref{fig. 7}(b) and (d), with a fixed $\beta$, we find that changes in steady-state infection density caused by $\lambda$ are greater than those caused by $\lambda^{*}$. Similar comparisons reveal that in terms of impact on steady-state infection density, $\lambda > \theta > \lambda^{*}$. In fact, in Figs. \ref{fig. 7}(b), (d), and (f), the average infection densities are around 0.329, 0.437, and 0.391, respectively, which is consistent with our observation. However, we also find that the changes in heatmaps for both awareness density and infection density are similar, and their maximum values are very close, indicating that $\lambda$, $\theta$, and $\lambda^{*}$ have a similar magnitude of importance and impact on epidemic spreading.

Regarding the impact of $\lambda$, $\theta$, $\lambda^{*}$, and integrated information on the density of aware nodes and infected nodes, we specifically illustrate these effects in Fig. \ref{fig. 8}. Observing Fig. \ref{fig. 8}(a), we find that, in terms of the impact on epidemic information diffusion, the relationship can be described as $\lambda > \theta \approx \lambda^{*}$. Similarly, in Fig. \ref{fig. 8}(b), regarding the effectiveness of curbing epidemic spreading, the relationship is $\lambda > \theta > \lambda^{*}$.

Overall, the integrated information, which includes three types of information propagation, results in the highest density of awareness, the highest infection threshold $\beta_{c}$, and the lowest density of infection. This is because a higher density of awareness means that more nodes are less susceptible to infection, leading to a lower overall infection density. In
contrast, as we disregard external information, the awareness density of the entire population is the lowest and the infection density is the highest. In other words, the diffusion of information indeed helps to suppress the spreading of the epidemic. Therefore, in real life, it is essential to maintain connections with friends and family, stay informed about relevant epidemic information in the media, and remain vigilant in the real world. This not only maximizes self-protection but also protects the family, society, and nation.

\subsection{Application of Information in the U.S. Power Grid}\label{sec: section 3.5}

To validate the applicability of our model to real-world cyber-physical systems, we further conducted simulations on a power grid system. The dataset comes from the U.S. power grid, with 4,941 nodes and 6,594 edges, and an average degree of 2.67 \cite{konect, r1}. In this system, power stations correspond to the nodes, and power lines correspond to the edges. Power stations fail at a rate of $\beta$, transitioning from the S state to the I state due to overload, short circuits, or attacks, and are repaired at a rate of $\mu$. Additionally, there is information communication between power stations via fiber optics or radio, where a station receives failure information from other stations at a rate of $\lambda$, and loses information at a rate of $\delta$. Power stations that receive failure information from other stations will adjust their behavior to prevent issues such as short circuits or overloads.

In the simulation of the power grid, we obtain similar results, which demonstrate the applicability of our model to real systems. Fig. \ref{fig. 9} illustrates the impact of various information on the affected density of the power grid and indicates that information propagation helps to increase the threshold that causes the outbreak of power station faults and reduces the extent to which the grid is affected. Among them, pairwise interaction information propagation is the most obvious for suppressing fault spreading, and 2-simplex and physical-layer information can also significantly reduce the size of faults. However, if we give reasonable and necessary consideration to all kinds of information, we can minimize the impact of faults and ensure the safety of power grids and the normal operation of social production and life.

\section{Conclusion and Outlook}\label{sec:section 4}

In conclusion, this work constructs a bidirectional coupled network of information propagation and epidemic spread, based on the coupled characteristics of entity interactions and information diffusion in cyber-physical systems. Using the epidemic model as an example, we integrate various types of information propagation, particularly the interaction between the physical layer and the cyber layer, focusing on the impact of physical-layer information on epidemic spread. We also explore the dynamic relationship between information diffusion and epidemic transmission. Additionally, we further validate our conclusions through simulations in the U.S. power grid. It is worth noting that, once the basic network structure and key parameter information are obtained, our model can easily simulate real-world scenarios, making it highly scalable and practically applicable, while also having a low computational complexity.

Our study demonstrates that in networked cyber-physical systems, there is a nonlinear relationship between infection rates and infection scales, or between power station failure rates and failure scales. This means that diseases with high infection rates or severe malicious attacks can lead to large-scale population infections or equipment malfunctions, triggering security issues and potentially causing system collapse. Our results also show that information propagation plays a highly effective and significant role in controlling infections or defending against malicious attacks. Specifically, in the case of epidemic spread, this is manifested in: Our research suggests that, first, pairwise interaction, as the most prominent form of information propagation, plays the most significant role in curbing epidemic spreading and increasing the density of the aware population. Second, 2-simplex information propagation, as a widespread channel of information propagation, not only raises the infection threshold and aware population density but also effectively reduces the infected population density. Notably, when individuals pay sufficient attention to epidemic information in the media and take it seriously, 2-simplex information propagation can even function similarly to pairwise interaction, greatly increasing the aware population density, thereby reducing the probability of infection and further suppressing the spreading of the epidemic. Finally, in our social activities, being vigilant and observant not only means giving enough care and attention to others, enhancing our understanding of each other but also providing us with relatively accurate information sources during an epidemic, reducing the infected population density overall. However, the information around us is often overlooked, it is more important than the information from the media since our social world is where we truly live, and it deserves more of our attention. Inspired by this work, we understand that in an information-rich region during an epidemic, the spread can be maximally contained by maintaining an observer’s mindset—proactively perceiving information and disseminating public opinion in time. The simulation results suggest that information indeed plays a crucial role in epidemic prevention and control or in mitigating power grid failures. 

However, we have overlooked other factors, such as the inevitable addition and elimination of nodes in real-world networks. To address these issues, we propose using queuing theory to simulate the population birth and death mechanisms in the system. By constructing a queuing model for the system, where the birth of individuals follows a Poisson process with an arrival rate of \(\nu\), and the time individuals spend in the network follows an exponential distribution with a rate parameter \(\mu\), we can theoretically simulate the birth and death of individuals in a real-world context. In addition, many real-world networks are not standard models like ER, WS, or BA networks, but are dynamic or weighted. We hope to find stable dynamic network algorithms and networks with reasonable weight distributions, in order to further explore the impact of information propagation on cyber-physical systems and validate and expand our conclusions in more real-world networks. Furthermore, the actual impact of objective behaviors, such as misjudgment by individuals when identifying infected nodes, on cyber-physical systems will be comprehensively and thoroughly explored in our future research. Finally, regarding model validation, although the effectiveness of the MMCA method has been proven, in order to draw more comprehensive and in-depth conclusions, using the PRISM tool for re-validation of the results will help strengthen the credibility of the findings, verify their general applicability, and contribute to advancing scientific research.

\bibliographystyle{IEEEtran} 


\begin{IEEEbiography}[{\includegraphics[width=1in,height=1.25in,clip,keepaspectratio]{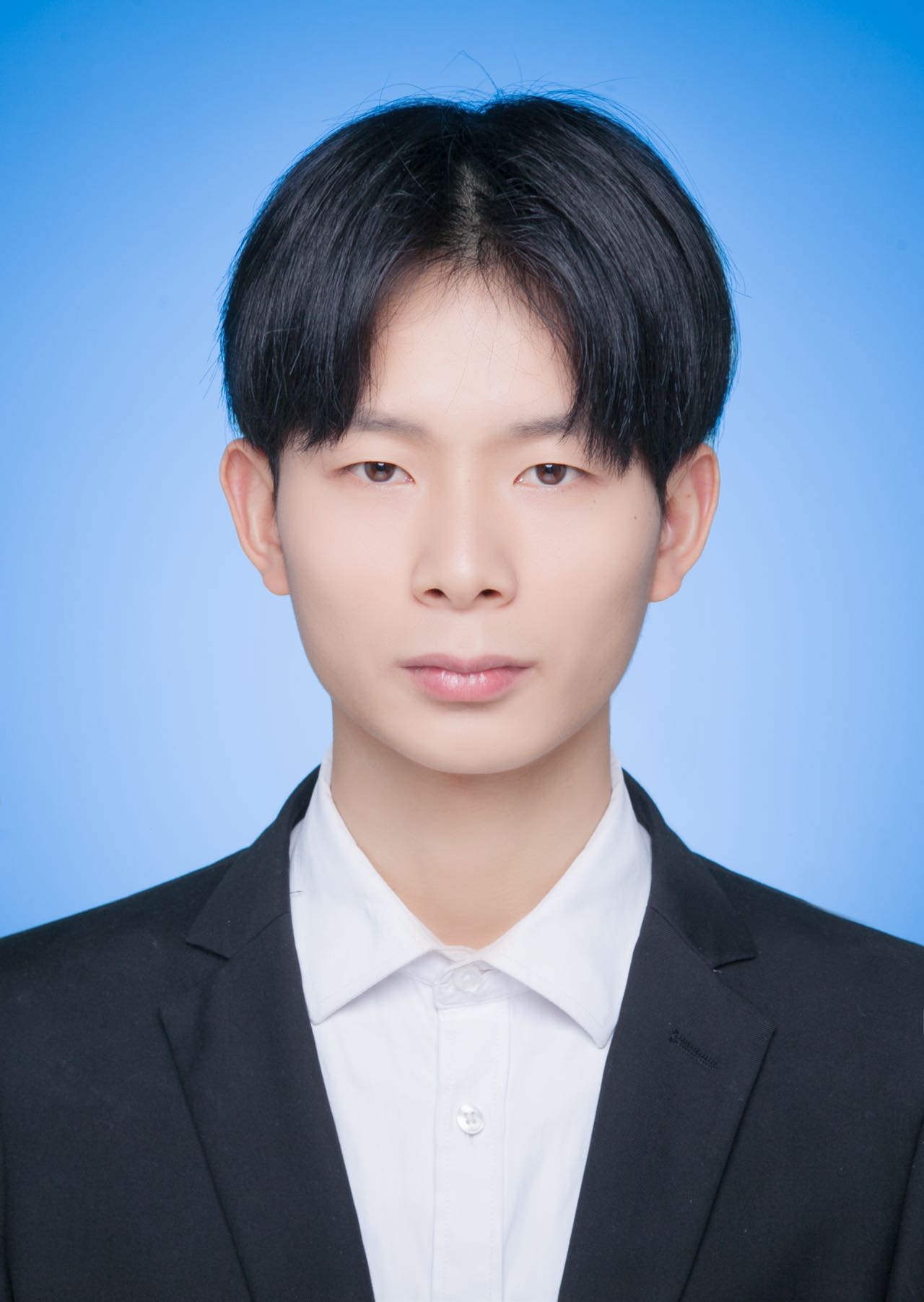}}]{Xianglai Yuan} received a bachelor’s degree from the School of Energy and Mechanical Engineering, Jiangxi University of Science and Technology, Jiangxi, China. He is currently pursuing a master’s degree in computer science and technology.

His research interests include Complex Networks, Network Propagation, Stochastic Processes, and Nonlinear Science.
\end{IEEEbiography}
\begin{IEEEbiography}[{\includegraphics[width=1in,height=1.25in,clip,keepaspectratio]{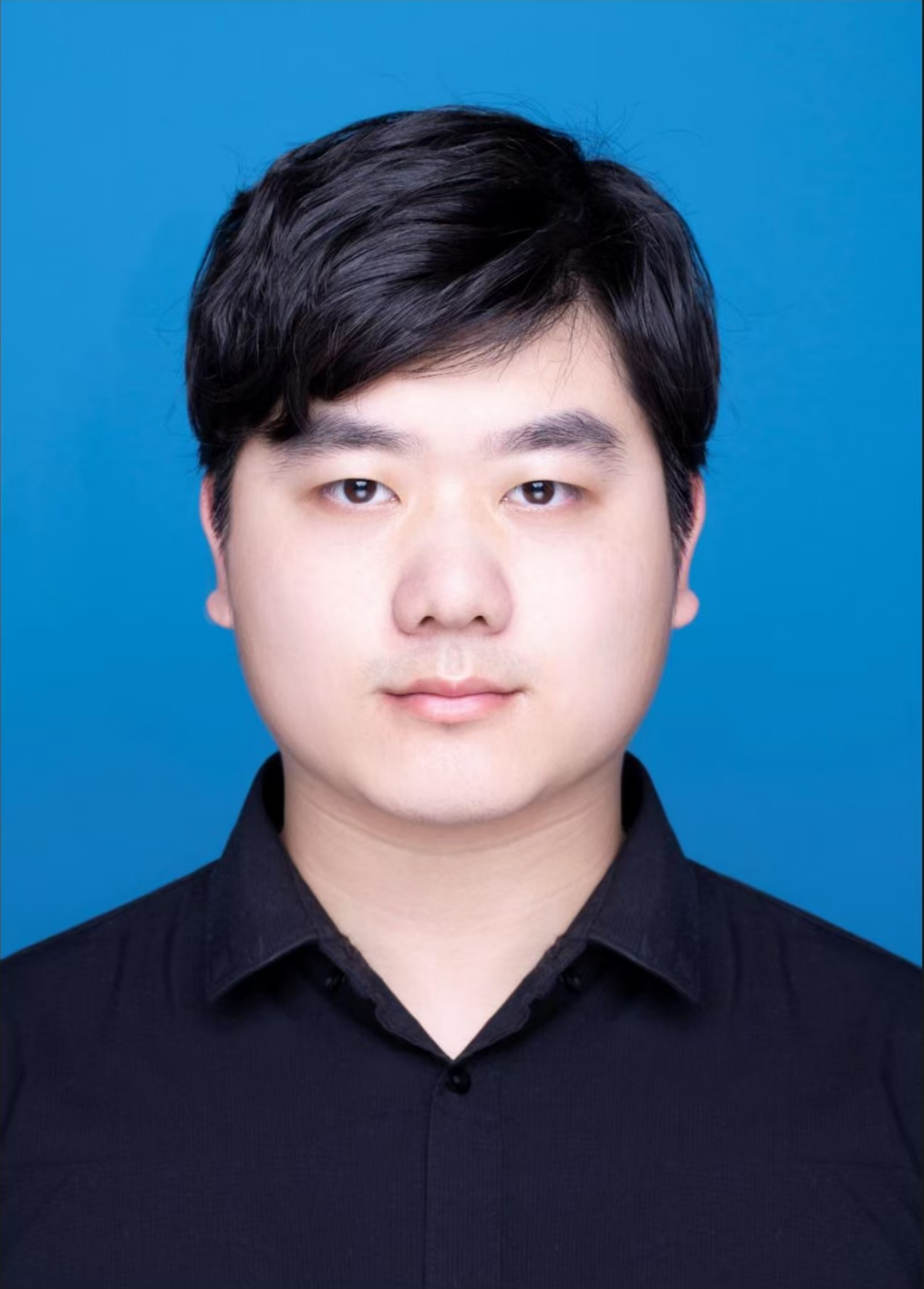}}]{Yichao Yao}
received a bachelor’s degree from the College of Artificial Intelligence, Southwest University, Chongqing, China. He is currently pursuing a master’s degree in computer science and technology. 

His research interests include Complex Networks, Evolutionary Games, Stochastic Processes, and Nonlinear Science.
\end{IEEEbiography}
\begin{IEEEbiography}[{\includegraphics[width=1in,height=1.25in,clip,keepaspectratio]{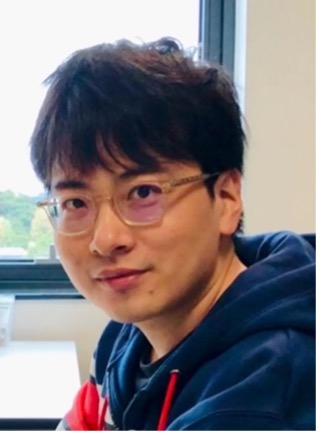}}]{Han Wu}
(Member, IEEE) is currently an Assistant Professor in the School of Electronics and Computer Science at University of Southampton, United Kingdom. He received his PhD in Computer Science from the Free University of Berlin, Germany. He has authored research papers in leading conferences, including USENIX Security, DSN, ISSRE, etc. He has served as a reviewer for IEEE Transactions on Information Forensics and Security, IEEE Transactions on Dependable and Secure Computing, etc.

His research interests include Federated Learning, Adversarial Machine Learning, Differential Privacy, and Cyber-Physical Systems.
\end{IEEEbiography}
\begin{IEEEbiography}[{\includegraphics[width=1in,height=1.25in,clip,keepaspectratio]{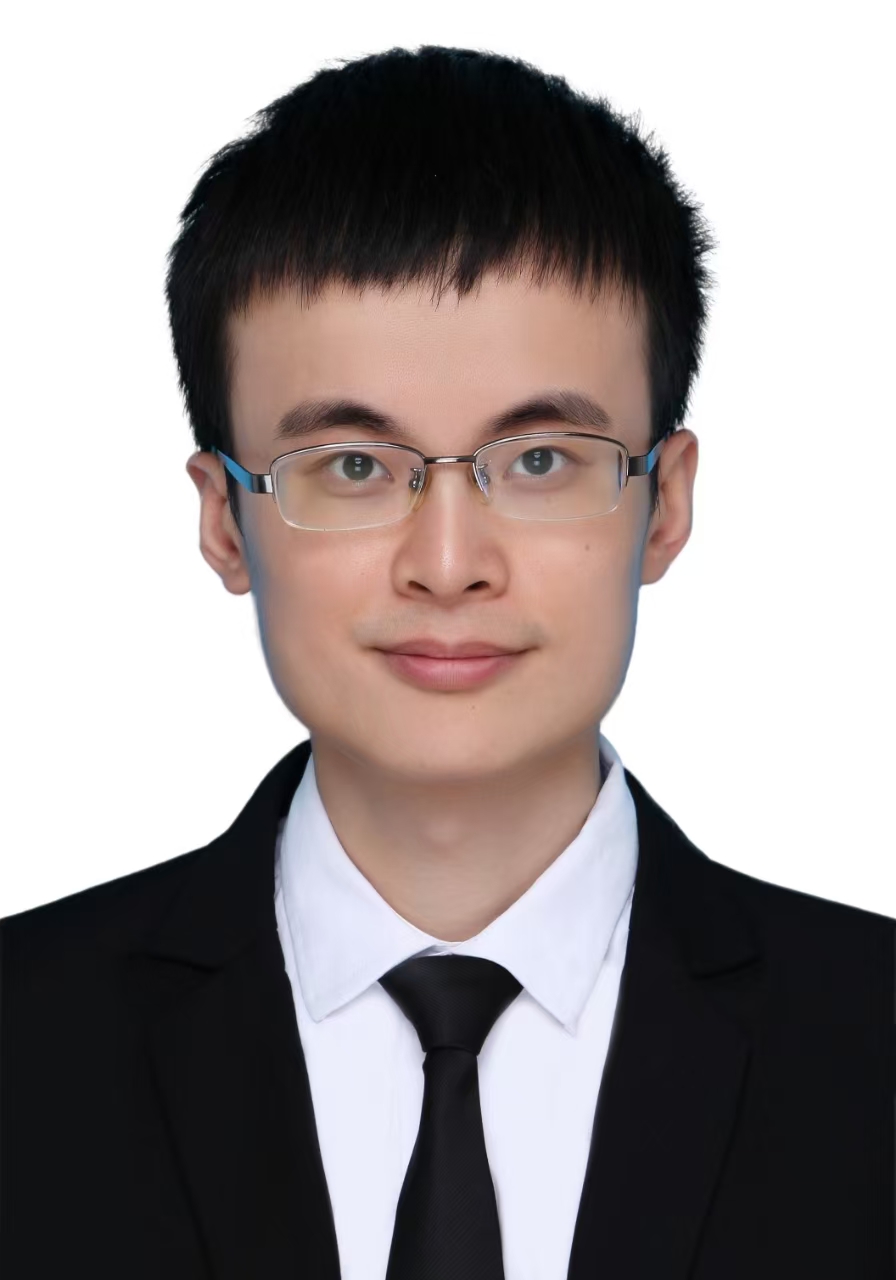}}]{Minyu Feng}
(Senior Member, IEEE) received his Ph.D. degree in Computer Science from a joint program between University of Electronic Science and Technology of China, Chengdu, China, and Humboldt University of Berlin, Berlin, Germany in 2018. Since 2019, he has been an associate professor at the College of Artificial Intelligence, Southwest University, Chongqing, China. 

Dr. Feng is a Senior Member of China Computer Federation (CCF) and Chinese Association of Automation (CAA). Currently, serves as a Subject Editor for Applied Mathematical Modelling, and an Editorial Board Member for Humanities \& Social Sciences Communications, Scientific Reports and International Journal of Mathematics for Industry. Besides, he is a Reviewer for Mathematical Reviews of the American Mathematical Society.

His research interests include Complex Systems, Evolutionary Game Theory, Computational Social Science, and Mathematical Epidemiology.
\end{IEEEbiography}

\end{document}